\documentclass[twocolumn,twocolappendix]{aastex631}

\newcommand{\rmd}{{\rm d}}

\usepackage{amsmath}
\usepackage{multirow}
\usepackage{bm}
\usepackage{subfigure}
\shorttitle{Modeling the impacts of galaxy IA on WL peak statisitcs}
\shortauthors{Zhang et al.}
\graphicspath{{./}{figures/}}

\begin{document}

\title{Modeling the impacts of galaxy intrinsic alignments on weak lensing peak statistics}

\correspondingauthor{X.K.Liu and Z.H.Fan}
\email{liuxk@ynu.edu.cn, zuhuifan@ynu.edu.cn}

\author{Tianyu Zhang}
\affiliation{South-Western Institute for Astronomy Research, Yunnan University, Kunming 650500, China}

\author{Xiangkun Liu*}
\affiliation{South-Western Institute for Astronomy Research, Yunnan University, Kunming 650500, China}

\author{Ziwei Li}
\affiliation{South-Western Institute for Astronomy Research, Yunnan University, Kunming 650500, China}

\author{Chengliang Wei}
\affiliation{Purple Mountain Observatory, Chinese Academy of Sciences, Nanjing 210023, China}

\author{Guoliang Li}
\affiliation{Purple Mountain Observatory, Chinese Academy of Sciences, Nanjing 210023, China}
\affiliation{Zhejiang University-Purple Mountain Observatory Joint Research Center for Astronomy, Zhejiang University, Hangzhou 310027, China}

\author{Yu Luo}
\affiliation{Department of Physics, School of Physics and Electronics, Hunan Normal University, Changsha 410081, China}

\author{Xi Kang}
\affiliation{Purple Mountain Observatory, Chinese Academy of Sciences, Nanjing 210023, China}
\affiliation{Institute for Astronomy, the School of Physics, Zhejiang University, Hangzhou 310027, China}

\author{Zuhui Fan*}
\affiliation{South-Western Institute for Astronomy Research, Yunnan University, Kunming 650500, China}



\begin{abstract}
Weak gravitational lensing (WL) peak statistics capture cosmic non-linear structures and can provide additional cosmological information complementary to cosmic shear two-point correlation analyses. They have been applied to different WL surveys successfully. To further facilitate their high precision applications, it is very timely to investigate the impacts of different systematics on WL peak statistics and how to mitigate them. Concerning the influence from galaxy intrinsic alignments (IAs), in this paper, we develop a theoretical model for WL high peaks taking into account the IA effects. It is an extension of our previous halo-based model. The IA corrections mainly include the modification of the lensing profile of clusters of galaxies due to the alignments of satellite galaxies and the additional shape noise correlations. We validate our model using simulations with the semi-analytical galaxy formation. We consider the cases where the satellite galaxies are averagely radially aligned toward the centers of their host clusters but with different dispersions $\sigma_{\theta}$. We show that our model works well for $\sigma_{\theta}>45^{\circ}$. If the IA corrections are not included in the model, for the Euclid/CSST-like source galaxy distribution and the survey area of $\sim 1000\deg^2$, the IA induced bias on $S_8$ can reach $\sim 8\sigma$ even for $\sigma_{\theta}=75^{\circ}$. With our model, not only the bias can be well mitigated, but also we can constrain the satellite IA to the level of {\bf $\sigma(\sigma_{\theta})\sim \pm 24^{\circ}$} simultaneously from WL high peak analyses alone using data from such a survey.

\end{abstract}



\section{Introduction} \label{sec:intro}

Arising from the gravitational light deflections by cosmic large-scale structures, weak lensing (WL) effects are a unique cosmological probe enabling us to map the large-scale matter distribution, which is
composed of mainly dark matter \citep[e.g.,][]{BS2001, FuFan2014, Kilbinger2015, Mandelbaum2018}. 

With the great success of different WL surveys, such as the Canada-France-Hawaii Telescope Lensing Survey \citep[CFHTLenS;][]{Heymans2012}, the Dark Energy Survey \citep[DES;][]{DES}, 
the Kilo-Degree Survey \citep[KiDS;][]{KiDS2015}, and the Hyper Suprime-Cam SSP Survey \citep[HSC;][]{HSC2018}, the new generation Stage IV surveys are right around the corner, 
including the ground-based Legacy Survey of Space and Time (LSST) of Rubin Observatory \citep{Ivezic2019}, and the space missions of {\it {Euclid}} \citep{Euclid, EuclidOverview},
Roman Space Telescope \citep{Roman2019} and the China Space Station Survey Telescope \citep[CSST;][]{CSST, LiuD2023}. 
These surveys will observe more than a billion source galaxies for WL studies, one to two orders of magnitude 
more than the currently available data.
They target for highly precise and accurate WL cosmological studies, therefore understanding and controlling different systematic errors will be critically important.   

To mitigate the impacts from systematics, one strategy is to downweight the mostly affected regions, such as the small-scale signals, in deriving cosmological constraints. This however inevitably leads to information loss. 
For the space-based Stage IV surveys, their high spatial resolutions can dramatically reduce the image blending effects \citep[e.g.,][]{EuclidOverview, LiuD2023}, giving rise to much improved shape measurements in dense
regions, such as in clusters of galaxies. To fully take the advantage of these surveys, we need to understand the possible systematic effects on small scales but not to simply ignore the signals there. 
Furthermore, with good modeling of the impacts from astrophysics-related systematics, we can potentially extract important astrophysical information while at the same time mitigating the biases on 
cosmological constraints.  

The galaxy intrinsic alignments (IAs) are closely related to the galaxy formation and evolution, carrying
valuable physical information \citep[e.g.,][]{Joachimi2015,Huang2018,Lan2024,Knebe2020,Tenneti2021}. 
On the other hand, they are one of the major systematics in WL cosmological studies that rely on measuring the galaxy shape correlations \citep[e.g.,][]{Troxel}.  
At the cosmic shear two-point correlation or power spectrum level, different models have been developed for the intrinsic-intrinsic (II) and gravitational shear-intrinsic (GI) signals
\citep[e.g.,][]{Bridle2007, HS04, Blazek2019}. They have been applied in WL observational analyses to control the IA-induced biases on cosmological parameter constraints
\citep[e.g.,][]{Abbott2022, Heymans2021, HSC2023}. 

As the peak statistics that capture the non-Gaussian features in lensing maps
are becoming a routine application in WL survey studies \citep[e.g.,][]{LiuJ2015, LiuX2015, Liu2016, Kacp, Shan2018, Martinet2018,Zurcher2022,Liu2023, Harnois2024,Marques2024}, investigations of the IA impacts on them are increasingly important and desired. At the moment, there is still not a concrete theoretical model for that yet. 
In \cite{Kacp} and \cite{Zurcher2022}, they estimate the satellite IA effects on peak SNRs and make corresponding corrections in their simulation-based templates to study the possible bias on cosmological constraints using DES data. 
In \cite{Harnois2022, Harnois2024}, they apply an infusion method to add the IA effects at the map pixel level based on the nonlinear tidal alignment model (NLA) \citep{Bridle2007} imposing a weight that is linearly proportional to 
the local surface matter density ($\delta$-NLA). The IAs are then interpolated to the positions of galaxies to construct their intrinsic ellipticities in building simulation mocks for cosmological inferences with KiDS and DES data.  
In \cite{Lu2023}, they use deep learning to extract cosmological information from HSC first-year data, where the IAs are constructed also using the NLA model. 
It is known that the NLA model cannot describe accurately satellite IAs in clusters of galaxies \citep[e.g.,][]{SB2010, Fortuna2021, Zhang2022}. Even with $\delta$-NLA, the peak analyses are limited to the SNR $\le \sim 4$ 
to avoid the very high SNR peaks that are dominantly originated from clusters of galaxies \citep{ Harnois2024}.

In \cite{Zhang2022}, we study the IA impacts on peak statistics systematically using numerical simulations with semi-analytical galaxy formation. We pay particularly attention to satellite IAs by varying their strength 
to analyze the dependence of the peak counts on satellite IAs. We find that the effects from satellite IAs are closely associated with their number boost in the source galaxy sample in cluster regions, which is also emphasized in \cite{Kacp}.
Given the boost information from our mock analyses, we show that the peak counts, in particular high peaks, are sensitive to the strength of satellite IAs. 
Besides, the IAs (central and satellite) contribute to shape noise correlations, altering the noise properties, which also affect the peak counts \citep[see also][]{Fan2007}. 

Inspired by the results from simulation analyses of \cite{Zhang2022}, in this paper, we present a theoretical model for WL high peak statistics taking into account the IA impacts. 
It bases on our previous studies of \cite{Fan2010} and \cite{Yuan2018}.

The paper is structured as follows. Sec.\ref{sec:theory} present the theoretical aspects related to the WL theory and the peak model ingredients. We describe in detail the needed corrections when the IAs are included.
In Sec.\ref{sec:modelperform}, with mock simulations, we demonstrate the procedures to build the IA corrected peak model and validate the model by comparing its predictions with the mock peak data. 
We also show how the model can mitigate the IA-induced bias in cosmological parameter constraints from WL high peak counts, and the potential to constrain the strength of satellite IAs together with the cosmological parameters.
Summary and discussions are presented in Sec.\ref{sec:discussion}.

\section{The IA-corrected model for WL high peak counts} \label{sec:theory}

WL high peaks are dominantly contributed by individual massive clusters of galaxies along lines of sight 
\citep[e.g.,][]{Hamana2004,Yang2011,Yuan2018, FLAMINGOpeak}. With this consideration, we have developed a halo-based model to predict the cosmological dependence 
of their abundances. The main physical ingredients of the model are the halo mass function for massive clusters, their density profiles 
and the cosmic distances \citep{Fan2010}. Additionally, we do the forward modeling of the impacts of the shape noise arising 
from galaxy intrinsic ellipticities and the projection effect from large-scale structures other than the massive clusters \citep{Yuan2018}. The model has been applied to 
different WL surveys successfully \citep{LiuX2015, Liu2016, Liu2023, Shan2018} and also extended to the peak steepness statistics \citep{LZW}. 

Regarding IAs, in the simulation analyses of \cite{Zhang2022}, we show that for high peaks, they are mainly affected by 
the IAs of the satellite galaxies of massive clusters in the source sample. They change the lensing profiles of their host clusters. 
In addition, the IAs induce physical correlations for the shape noise. Within our halo-based modeling approach, these IA impacts can be 
incorporated relatively straightforwardly. 

In this section, we first summarize the basic WL theory and our established halo-based model for WL high peaks. We then describe the IA-corrected model in detail.

\subsection{Brief Theory of Weak Lensing Effects} \label{sec:WL}

The cosmic large-scale structures gravitationally bend the light propagations, resulting in changes of the observed images of distant galaxies in comparison with their intrinsic ones. 
In the WL lensing limit, the distorsions are weak, and can be characterized by two physical quantities, the convergence $\kappa$ and the shear $\bm{\gamma}=\gamma_1+i\gamma_2$. 
The convergence leads to an isotropic size change and the shear gives rise to anisotropic shape distortions of the observed galaxy images. They depends on the lensing potential $\psi$.
Concretely, the WL effects can be described by the Jacobian matrix $A$ \citep{BS2001}:

\begin{equation}
\label{eq:Jacobmatrix}
    A=\bigg(\delta_{ij}-\frac{\partial^2\psi(\boldsymbol{\theta})}{\partial\theta_i \partial\theta_j}
        \bigg)
    =\begin{pmatrix}
    1-\kappa-\gamma_1 & -\gamma_2\\
    -\gamma_2 & 1-\kappa+\gamma_1
    \end{pmatrix},
\end{equation}
where $\boldsymbol{\theta}=(\theta_1,\theta_2)$ is the angular position on the sky. The $\kappa$ and $\bm{\gamma}$ are mutually dependent with the following relation in the Fourier space:  

\begin{equation}
\label{eq:kap2gam}
    \tilde {\boldsymbol \gamma}(\boldsymbol{k})=\frac{1}{\pi}\tilde{\boldsymbol D}(\boldsymbol k)\tilde{\kappa}(\boldsymbol k),
\end{equation}
with
\begin{equation}
\label{eq:D}
    \tilde{\boldsymbol D}(\boldsymbol{k})=\pi\frac{k_1^2-k_2^2+2ik_1k_2}{k_1^2+k_2^2}.
\end{equation}

Under the first order Born approximation, $\kappa$ can be formulated by
\begin{equation}
\label{eq:kappa}
\begin{split}
    \kappa(\boldsymbol{\theta})=&\frac{3H_0^2\Omega_{\rm m}}{2c^2}\int_0^{\chi_H} d\chi'\int^{\chi_H}_{\chi'}d\chi \\
&\bigg [p_s(\chi)\frac{f_K(\chi-\chi')f_K(\chi')}{f_K(\chi)a(\chi')}\bigg ]{\delta[f_K(\chi')\boldsymbol{\theta},\chi']},
\end{split}
\end{equation}
where $\chi$, $f_K$ and $a$ are the comoving radial distance, comoving angular diameter distance and the cosmic scale factor, respectively. The quantity $\chi_H$ is the comoving horizon with
$\chi_H=\chi(z=\infty)$. The function $p_s$ is the source galaxy distribution function and $\delta$ is the 3D matter density fluctuation. Therefore $\kappa$ is the dimensionless projected 
surface matter density weighted by the lensing efficiency kernel and the source distribution function. 

In WL observations, the direct observables are the galaxy ellipticities measured from their light distributions. They consist of both the intrinsic ellipticities $\boldsymbol{\epsilon_s}$ and the
WL shears. Specifically \citep{Seitz97},  
\begin{equation}\label{eq:epsilon2g}
    \boldsymbol \epsilon= 
    \begin{cases}
        \dfrac{\boldsymbol \epsilon_s+\boldsymbol g}{1+\boldsymbol g^{*} \boldsymbol \epsilon_s}, & |g| \leq 1 \\ 
        \dfrac{1+\boldsymbol g \boldsymbol \epsilon_s^*}{{\boldsymbol \epsilon_s}^*+\boldsymbol g^{*}}, & |g|>1
    \end{cases},
\end{equation}
where $\bm{g}$ is the reduced shear with $\bm{g}=\bm{\gamma}/(1-\kappa)$. Without considering IAs, we have $\langle{\bm{\epsilon}}\rangle=\langle\bm{g}\rangle$ 
when averaging over infinite number of galaxies. For a finite sample, there is a left-over residual shape noise from galaxy intrinsic ellipticities. 
With IAs, they contribute to the average, and thus contaminate the WL fields and subsequently the peak statistics.

In our analyses here, we focus on peaks from WL convergence fields. For model comparisons with simulations in Sec.{\ref{sec:modelperform}}, we build mock observed  
galaxy samples, and apply the nonlinear Kaiser-Squires method to reconstruct the convergence fields from ellipticities using Eq.\eqref{eq:kap2gam} \citep{Kaiser93,Seitz95,KS96}. 

For that, we first smooth the mock galaxy ellipticities onto a grid to obtain the shear maps $\langle{\boldsymbol{\epsilon}}\rangle(\boldsymbol \theta)$ by

\begin{equation}
\label{eq:smooth}
    \langle{\boldsymbol {\epsilon}}\rangle(\boldsymbol{\theta})=\frac{\Sigma_{j}W_{{\theta}_{\rm G}}(\boldsymbol{\theta}_j-\boldsymbol{\theta}){\boldsymbol {\epsilon}}(\boldsymbol{\theta}_j)}{\Sigma_{j}W_{{\theta}_{\rm G}}(\boldsymbol{\theta}_{j}-\boldsymbol{\theta})},
\end{equation}
where $\boldsymbol \theta_j$ is the position of a galaxy, and $W$ is the smoothing function. 
Here we adopt the Gaussian smoothing function,
\begin{equation}
    \label{eq:Gausswindow}
        W_{\theta_{\rm G}}(\boldsymbol{\theta})=\frac{1}{\pi\theta_{\rm G}^2}\rm{exp}(-\frac{|\boldsymbol{\theta|}^2}{\theta_{\rm G}^2}),
\end{equation}
where $\theta_{\rm G}$ is the smoothing scale.

We follow \cite{LiuX2015} to reconstruct the convergence fields from the reduced shear iteratively. At the beginning of the iterations, $\kappa^{(0)}=0$ and ${\bm \gamma}^{(0)}=\langle{{\bm \epsilon}}\rangle$. 
Then $\kappa^{(n)}$ is calculated by Eq.\eqref{eq:kap2gam} from ${\bm \gamma}^{(n-1)}$, and ${\bm \gamma}^{(n)}$ is updated to ${\bm \gamma}^{(n)}=(1-\kappa^{(n)})\langle{{\bm \epsilon}}\rangle$ in each iteration. 
This ends when the difference of convergence maps $(\kappa^{(n)}-\kappa^{(n-1)})$ reaches the accuracy limit, set as $10^{-6}$ in our analyses. The final $\kappa^{(n)}$ is 
the reconstructed convergence field, denoted by $K_N$, for our peak analyses.

\subsection{Summary of the halo-based model for high peaks} \label{sec:F10}

By assuming that WL high peaks arise mainly from individual massive clusters of galaxies along lines of sight, we have established a theoretical framework to calculate high peak abundances 
taking into account the impacts from the shape noise and the large-scale structure projection effects. The details can be found in \cite{Fan2010} and \cite{Yuan2018}. Here we
present a summary of the model. 

For physical considerations, we write the convergence field as 

\begin{equation}
\label{eq:KN}
K_{N}=K_{\rm{H}}+K_{\rm{LSS}}+N.
\end{equation}
Here $K_{\rm{H}}$ represents the part from massive clusters of galaxies with mass $M\ge M_*$, 
where $M_*$ is in the order of $10^{14}h^{-1}\hbox{M}_{\odot}$ as suggested by simulation analyses \citep{Yuan2018, Wei2018Halo,FLAMINGOpeak}. The quantity $K_{\rm{LSS}}$ is the contribution from large-scale structures not included in $K_{\rm{H}}$, and $N$ is the residual shape noise. 
For $K_{\rm{H}}$, each massive cluster is modeled using the Navarro-Frenk-White density profile \citep[NFW,][]{NFW1,NFW2}. For $K_{\rm{LSS}}$ that contains the cumulative contributions from halos with $M<M_*$
and the halo-halo correlations, we approximate it as a Gaussian random field \citep{Yuan2018}. The noise field $N$ is also taken to be Gaussian. Thus $K_{N}$ can be regarded as a Gaussian random field
modulated by the heavily non-Gaussian structures of massive clusters. The peak abundances can then be calculated using the Gaussian random field theory \citep{BBKS, BE1987}.

We divide the considered areas into halo regions occupied by massive clusters of $M\ge M_*$, and the field regions outside these halos. The total peak number density distribution can be written as
\begin{equation}
\label{eq:peaknum}
    n_{\rm peak}(\nu)\rmd \nu = n_{\rm peak}^c(\nu)\rmd \nu + n_{\rm peak}^n(\nu)\rmd \nu,
\end{equation}
where $n_{\rm peak}^c(\nu)$ and $n_{\rm peak}^n(\nu)$ are the peak number densities in halo and field regions, respectively. Here $\nu=K_N/\sigma_0$ is the signal-to-noise-ratio (SNR) of a peak, 
and $\sigma_0$ is the rms of the total Gaussian random field $K_{\rm{LSS}}+N$. 

By definition, a peak in the $K_N$ field has the first derivatives $K_N^i=\partial_i K_N=0$ and the second derivatives $K_N^{ij}=\partial_i\partial_j K_N$ negative finite. 
For the peak number density in halo regions, from the Gaussian random field theory taking into account the modulations from $K_{\rm{H}}$, we have \citep{Fan2010,Yuan2018}

\begin{equation}
    \label{eq:npeak_halo}
    \begin{split}
        n_{\rm peak}^c(\nu)&=\int\rmd z \frac{\rmd V(z)}{\rmd z \rmd \Omega}\int_{M_*} \rmd M \frac{\rmd n(M,z)}{\rmd M}\\
        &\times \int_0^{\theta_{\rm vir}} \rmd \theta (2\pi\theta)\hat{n}^c_{\rm peak}(\nu,\theta,M,z),
    \end{split}
\end{equation}
where $\rmd n(M,z)/ \rmd M$ is the halo mass function (HMF), $\theta_{\rm vir}$ is the angular virial radius of a halo, $\rmd V(z)$ and $\rmd \Omega$ are the unit volume at redshift $z$ 
and the unit solid angle, respectively. The quantity $\hat{n}_{\rm peak}^c (\nu, \theta,M,z)$ in the integration is given by 

\begin{equation}
    \label{eq:hatnpeak_halo}
        \begin{split}
        &\hat{n}_{\rm peak}^c (\nu,\theta,M,z)=
        \exp\left[ -\frac{(K_{\rm{H}}^1)^2+(K_{\rm{H}}^2)^2}{\sigma_1^2} \right]
        \left[ \frac{1}{2\pi\theta_*^2} \frac{1}{(2\pi)^{1/2}} \right]\\
        &\times
        \exp \left[ -\frac{1}{2} \left( \nu-\frac{K_{\rm{H}}}{\sigma_0} \right)^2 \right]
        \times \int_0^\infty \rmd x_{}
        \left\{
        \frac{1}{[2\pi(1-\gamma^2)]^{1/2}}\right.\\
        &\times
        \exp \left[ -\frac{[x+(K_{\rm{H}}^{11}+K_{\rm{H}}^{22})/\sigma_2-\gamma(\nu-K_N/\sigma_0)]^2}{2(1-\gamma^2)} \right]
        \\
        &\times F(x) \},
        \end{split}
\end{equation}
where $\theta_*=2\sigma_1^2/\sigma_2^2$, $\gamma=\sigma_1^2/(\sigma_0\sigma_2)$, and
\begin{equation}
    \label{eq:Ffunction}
    \begin{split}
        F(x)&=
        \exp \left[ -\frac{(K_{\rm{H}}^{11}-K_{\rm{H}}^{22})^2}{\sigma_2^2} \right]\\
        &\times
        \int_0^{1/2} \rmd e \hbox{ } 8(x^2 e) x^2 (1-4e^2) \exp (-4x^2 e^2)\\
        &\times
        \int_0^\pi \frac{\rmd \theta}{\pi} \exp \left[ -4 x e \cos(2\theta) \frac{K_{\rm{H}}^{11}-K_{\rm{H}}^{22}}{\sigma_2}\right].
    \end{split}
\end{equation}
Here $x=-(K_N^{11}+K_N^{22})/\sigma_2$. The quantities $\sigma^2_i$ ($i=0,1,2$) are the moments of the total Gaussian random field $K_{\rm{LSS}}+N$:
\begin{equation}
    \label{eq:sigPS}
    \sigma_i^2=\sigma_{{\rm LSS},i}^2+\sigma_{{\rm N},i}^2=\int_{0}^{\infty} \frac{\ell \rmd \ell}{2\pi} \ell^{2i} (C_\ell^{{\rm LSS}}+C_\ell^{{\rm N}}),
\end{equation}
where $C_\ell^{\rm{LSS}}$ and $C_\ell^{\rm{N}}$ are the power spectrum of $K_{\rm{LSS}}$ and $N$, respectively. 

For the shape noise field $N$, with the Gaussian smoothing function Eq.\eqref{eq:Gausswindow}, we have  
\begin{equation}
    \label{eq:sig012}
    \sigma^2_{\rm N,0}=\frac{\sigma_{\epsilon}^2}{4 \pi n_{\rm g} \theta_{\rm G}^2},
\end{equation}
and $\sigma_{\rm N,0}:\sigma_{\rm N,1}:\sigma_{\rm N,2}=1:\sqrt{2}/\theta_{\rm G}:2\sqrt{2}\theta_{\rm G}^2$. The quantities $\sigma_{\epsilon}$ and $n_{\rm g}$ are the dispersion of 
the total galaxy intrinsic ellipticities and the number density of source galaxies, respectively.

For $K_{\rm{LSS}}$, its power spectrum contains the one-halo contributions from halos with $M<M_*$ and the two-halo terms of all the halos. In our modeling, its corresponding 3D matter power spectrum is
calculated by \citep{Yuan2018}
\begin{equation}
    \label{eq:P_LSS}
        P_\delta^{\rm{LSS}}=P_\delta-P_\delta^{\rm{1H}}|_{M\geq M_*},
\end{equation}
where $P_\delta$ is the total nonlinear matter power spectrum, and $P_\delta^{\rm{1H}}|_{M\geq M_*}$ is the one-halo part from $M\ge M_*$ halos. With $P_\delta^{\rm{LSS}}$, 
we can obtain $C_{\ell}^{\rm{LSS}}$ by doing the integration with the lensing efficiency kernel to calculate $\sigma_{{\rm LSS},i}^2$. 

For the field regions outside massive halos, the peak number density can be calculated directly from the Gaussian random field $K_{\rm{LSS}}+N$. It is given by 
\begin{equation}
\label{eq:field}
    \begin{split}
    n_{\rm peak}^n(\nu)&=\frac{1}{\rmd \Omega}\left\{
    n_{\rm ran}(\nu)\left[ \rmd \Omega-\int \rmd z \frac{\rmd V(z)}{\rmd z} \right. \right.\\
    &\times\left. \left.  \int_{M_{\rm lim}} \rmd M n(M,z)(\pi\theta^2_{\rm vir}) \right]
    \right\},
    \end{split}
\end{equation}
where $n_{\rm ran}$ can be calculated from Eq.\eqref{eq:hatnpeak_halo} by setting $K,K^i,K^{ij}=0$.

We note that our model is valid for WL high peaks, where $K_{\rm{LSS}}$ contributions are much less than that from $K_{\rm{H}}$. In this case, the inaccuracy of the Gaussian assumption for $K_{\rm{LSS}}$ 
does not affect our calculations significantly.

\subsection{Including IAs in the model} \label{sec:F10mod}

In \cite{Zhang2022}, we show that for a relatively wide source redshift distribution, the IAs of the satellite galaxies in massive clusters affect WL high peak counts the most.
This is also addressed in e.g., \cite{Kacp,Harnois2022,Harnois2024}. It arises because of the high number density of galaxies in cluster regions. A significant fraction of satellite galaxies
are included in the source sample, causing a local boost of the souce galaxies there. The effects on the WL signals of their host clusters can be schematically written as follows

\begin{equation}\label{eq:IAdilute}
\langle{\bm\gamma}\rangle\sim \frac{\Sigma_{N_b} {\bm \gamma}+\Sigma_{N_c} \bm{\epsilon_s}}{N_b+N_c}
=\frac{\Sigma_{N_b} {\bm \gamma}}{N_b+N_c}\bigg(1+\frac{\Sigma_{N_c} \bm \epsilon_s}{\Sigma_{N_b} {\bm \gamma}}\bigg),
\end{equation}
where $N_b$ and $N_c$, respectively, are the number of background galaxies and the number of satellite member galaxies in the source sample in a cluster region, 
and $\bm{\epsilon_s}$ is the intrinsic ellipticities of satellite galaxies. If the satellite galaxies have no IAs, they lead to the dilution effect on the WL signal with 
$\langle{\bm\gamma}\rangle\sim ({N_b})/(N_b+N_c)\bm\gamma=\alpha_{\rm{dilution}}\bm\gamma$. With satellite IAs, an additional effect occurs as illustrated in the term
$\alpha_{\rm{IA}}=(1+\Sigma_{N_c} {\bm \epsilon_s}/\Sigma_{N_b} \bm{\gamma})$.

In terms of our model modifications, the above dilution and IA effects change the $K_{\rm{H}}$ related massive halo parts from the NFW profile to the one with the corrections from 
$\alpha_{\rm {dilution}}$ and $\alpha_{\rm{IA}}$. More specifically, 
\begin{equation}
\label{eq:KHchange}
K_{\rm{H}}=K_{\rm{H}}^{\rm{NFW}}\cdot \alpha_{\rm {dilution}} \cdot \alpha_{\rm{IA}}.
\end{equation}

In observational analyses, the boost of the number of satellite galaxies in cluster regions needs to be measured from the correlations of the source galaxy sample with the known cluster
catalogs. It can change with the mass and the redshift of the clusters. This has been applied in our previous studies of \cite{Shan2018} and \cite{Liu2023}
to estimate the dilution effects on peak analyses using KiDS-450 data and HSC-SSP S16A data, respectively. 
With the boost information, we can perform single-halo simulations with different levels of satellite IAs to obtain 
$\alpha_{\rm{IA}}$ for the corrections on $K_{\rm{H}}$ and the related first and second derivatives. 
The procedure will be shown explicitly in the next section with our mock simulation data. 

Besides the effects on $K_{\rm{H}}$ related terms, the boost and IAs also change the noise fields. The boost effect leads to the lower noise levels in cluster regions because of the higher 
number densities there. For that, we have developed a correction method in our model calculations \citep{Shan2018, Liu2023}. For cluster regions, we use the measured boost information
to calculate the local noise levels. For the field regions outside massive clusters, we estimate the average number density of source galaxies $n_g^{\rm{field}}$ for the noise calculations
by  
\begin{equation}
    \label{eq:boostnumber}
        n_{\rm g}S_{\rm eff}=\sum n_{\rm g}^{\rm halo} S_{\rm eff}^{\rm halo} + n_{\rm g}^{\rm field} S_{\rm eff}^{\rm field},
\end{equation}
where $n_g$ is the average number density of source galaxies over the total effective survey areas $S_{\rm eff}$, $n_g^{\rm{halo}}$ and $S_{\rm eff}^{\rm halo}$ are the 
local number density and the area occupied by a massive cluster with a given mass and redshift calculated using its angular virial radius, 
and $S_{\rm eff}^{\rm field}=S_{\rm eff}-\sum S_{\rm eff}^{\rm halo}$. The summation is over all the massive clusters. 

With IAs, the noise fields are further changed as IAs contribute to physical correlations of the shape noise \citep[e.g,][]{Fan2007}. 
Similar to our separations of $K_{\rm{H}}$ and $K_{\rm{LSS}}$, we divide the IA induced noise correlations into two parts. One is due to the
satellite IAs in massive clusters of galaxies, and the other part is from large-scale II and GI shape correlations 
adopting the NLA model \citep{HS04,Bridle2007,Joachimi2011}.
This consideration is well justified by the studies of, e.g., \cite{SB2010} and \cite{Zhang2022}. They show that while the NLA model 
can describe the large-scale II and GI well, the one-halo small-scale satellite IAs needs to be treated separately. In our analyses here, for the satellite IA part, 
the total noise including the random shape noise plus the satellite-satellite II and the background-satellite GI contributions
can be calculated from the variance of different realizations of our large number single-halo simulations, noted as $\sigma^2_{{\rm map},i} ({\rm halo})=\sigma^2_{{\rm N},i}({\rm halo})+\sigma^2_{{\rm IA},i}({\rm satellite})$, which will be detailed in the next section.

For the large-scale II and GI contributions, the corresponding power spectra from NLA are given by 
\begin{equation}
\label{eq:II}
    P_{\rm II}(k,z)=F_{\rm IA}^2(z)P_\delta^{\rm NL}(k,z),
\end{equation}
\begin{equation} 
\label{eq:GI}
    P_{\rm GI}(k,z)=F_{\rm IA}(z)P_\delta^{\rm NL}(k,z),
\end{equation}
\begin{equation} 
\label{eq:Fz}
    F_{\rm IA}(z)=-A_{\rm IA}C_1\rho_{\rm crit}\frac{\Omega_{\rm m}}{D_+(z)}(\frac{1+z}{1+z_0})^{\eta_{\rm eff}},
\end{equation}
where $A_{\rm IA}$ is the amplitude parameter, $\rho_{\rm crit}$ is the current critical density and $D_+(z)$ is the linear growth factor. 
We take the fiducial $C_1=5\times10^{-14}h^{-2}M^{-1}_\odot{\rm Mpc}^3$ and $\eta_{\rm eff}=0$ in our calculations \citep{Wei}. 
From the power spectra, we can then calculate $C^{\rm II}_\ell+C^{\rm GI}_\ell$ and further their contributions to the noise parameter $\sigma^2_{{\rm IA},i}$ ($i=0,1,2$).

The total corrections for the noise parameters can be summarized as follows
\begin{equation}
    \label{eq:sigmatotal}
    \sigma_{{\rm total},i}^2=
    \begin{cases}
        \sigma^2_{{\rm map},i} ({\rm halo})+\sigma_{{\rm LSS},i}^2+\sigma^2_{{\rm IA},i}({\rm NLA})  \\
        \sigma^2_{{\rm N},i}({\rm field})+\sigma_{{\rm LSS},i}^2+\sigma^2_{{\rm IA},i}({\rm NLA})  
    \end{cases},
\end{equation}
where the first equation is for the massive cluster regions, and the second one is for the field regions. 

With the change on $K_{\rm{H}}$ from Eq.\eqref{eq:KHchange} and the noise change of Eq.\eqref{eq:sigmatotal}, we implement them into Eq.\eqref{eq:hatnpeak_halo} and Eq.\eqref{eq:field}
to calculate the IA-corrected WL high peak abundances.   

To be clear, we summarize the modeling procedures in the following steps.

\noindent {{\it {Step 1}}: Measuring the boost information of satellite galaxies in massive clusters of different redshift and mass bins from the known cluster catalogs in the survey areas. 
This needs to be done on the base of a specific survey.}

\noindent {{\it {Step 2}}: Performing single-halo simulations for different redshift and mass bins according to the boost information from Step 1 and considering different values of the satellite IA dispersion
parameter $\sigma_{\theta}$. We then obtain the correction term $\alpha_{\rm{dilution}}\alpha_{\rm{IA}}$
for $K_{\rm{H}}$, and also the satellite IA contributions to the noise field in the cluster regions $\sigma^2_{{\rm map},i} ({\rm halo})$.}

\noindent {{\it {Step 3}}: Including the large-scale II and GI contributions $\sigma^2_{{\rm IA},i}({\rm NLA})$ to the noise in 
the cluster and field regions, and calculating correctly $\sigma^2_{{\rm N},i}(\rm{field})$.}

\noindent {{\it {Step 4}}: With the corrections from Step 2 and Step 3, we calculate the model prediction for WL high peaks using Eqs.\eqref{eq:peaknum}, \eqref{eq:npeak_halo}, \eqref{eq:hatnpeak_halo} and \eqref{eq:field}.}

We note that only the information from Step 1 needs to be measured from the observational data of the considered survey. With that, we do the forwarding modeling through Step 2 to Step 4 
to include the IA effects in the theoretical prediction without doing any corrections to the data themselves.

\section{Model performance tests}\label{sec:modelperform}

In this section, we validate our IA-corrected model for WL high peaks using mock simulation data. 

\subsection{Mock simulations} \label{sec:mock}

The simulation data used here are similar to the ones used in \cite{Zhang2022}. They are generated from the full-sky ray-tracing simulation of \cite{Wei} with the base N-body simulations from ELUCID project
\citep{Wang2014, Wang2016}. The ray-tracing is done up to redshift $z\approx2.1$. The galaxy formation and evolution are included taking the semi-analytical approach \citep{Luo2016,FuJ2013,Guo2013}. 
For central galaxies, the orientations are assigned following their host dark matter halo mass distributions for elliptical galaxies and according to the halo spins for disk galaxies assuming the ratio of the thickness to the diameter of the disk is $r_d=0.25$ \citep{Joachimi2013,Wei}.  
For satellite disk galaxies, they are treated in the same way as that for centrals.
For elliptical satellites, their 3-D axial ratios are assigned according to a probability distribution derived from the 
identified halos in the simulation \citep{Wei}. To be described shortly, we change the correlation of the spin or the long-axis orientation 
of the disk/elliptical satellite galaxies for different level of IAs.  

From the galaxies in the simulation, we build a mock source sample according to the Euclid WL source redshift distribution \citep{Harnois2022, EuclidOverview}, which is very similar to the
one expected from CSST \citep{CSST,LiuD2023}. We take the number density of galaxies to be $n_g=30\hbox{ arcmin}^{-2}$.

We follow the procedure of \cite{Zhang2022} to assign IA signals to the satellite galaxies. The 3D orientation of a disk satellite is
taken to be perpendicular to its spin direction, and for an elliptical one, it is the direction of the long axis.
It is set in the local spherical coordinate system 
$(\theta_{\rm 3D},\phi)$ where $\theta_{\rm 3D}$ is the zenith angle between the satellite orientation direction and the line connecting it to its central galaxy. 
Taking into account an orientation angle dispersion $\sigma_\theta$, we adopt the following probability distribution for the satellite orientations:
\begin{equation}
    f(\theta_{\rm{3D}})\rmd\theta_{\rm{3D}}\rmd\phi=A\cdot\rm{exp}[-\frac{1}{2}(\frac{\theta_{\rm{3D}}-\theta_{0}}{\sigma_{\theta}})^2]\rmd\theta_{\rm{3D}}
    \rm{sin}(\theta_{\rm{3D}})\rmd\phi,
    \label{eq:theta_dis}
\end{equation}
where $A$ is the normalization parameter. 
The galaxies with 3D alignments are then projected to get their 2D intrinsic ellipticities $\bm \epsilon_s$, using the same approach shown in \citet{Joachimi2013} and \cite{Wei}. A more detailed description is given in Appendix \ref{app:IAsetting}.
The dispersion of the total $\bm \epsilon_s$ is $\sigma_{\epsilon_s}\sim 0.27$ from the simulated galaxies.

In our analyses, we set $\theta_0=0$, i.e., assuming that on average the satellite galaxies are radially aligned with a dispersion of $\sigma_\theta$.
When $\sigma_\theta$ goes to infinity, this distribution is back to the 3D uniform distribution, i.e., the satellite galaxies have no IAs. 
With $\sigma_\theta=0$, the satellite galaxies are perfectly radially aligned. Here we build mocks with different $\sigma_\theta$ with the larger value resulting in  
weaker IA signals. Specifically, we consider cases with $\sigma_{\theta}=75^{\circ}$, $60^{\circ}$ and $45^{\circ}$. We also build a set of mock data
with random satellite orientations, i.e., no IAs. In this case, the galaxy number boost still exists.

With the mock galaxy samples and their lensing signals from ray-tracing calculations, we generate the \textquotesingle observed\textquotesingle\hbox{ } ellipticity for each galaxy 
using Eq.\eqref{eq:epsilon2g}. 
For each sample with a given $\sigma_{\theta}$, we build $156$ patches of $3\times3\ {\rm deg}^2$ galaxy ellipticity data, and do the reconstruction to obtain $156$ convergence maps. 
The smoothing scale is $\theta_{\rm G}=2\hbox{ arcmin}$. For each map, the convergence field is sampled on $1024\times 1024$ pixels. For peak analyses, 
we exclude $5\times\theta_{\rm G}\sim57$ pixels from each edge of a map to suppress the boundary effects. The effective area for each sample is $\approx 1108.8\ {\rm deg}^2$.

\subsection{Measuring the number boost of satellite galaxies}\label{sec:boost}

As presented in Sec.\ref{sec:F10mod}, the satellite IA effects on WL high peaks are due to the excess number of satellite galaxies in the source sample. This number boost
can depend on survey conditions, and thus needs to be measured from data. In real observations, this can be obtained by the correlations of the source galaxy distribution 
with the known cluster positions in the survey areas \citep{Shan2018, Liu2023}. 

\begin{figure}
    \plotone{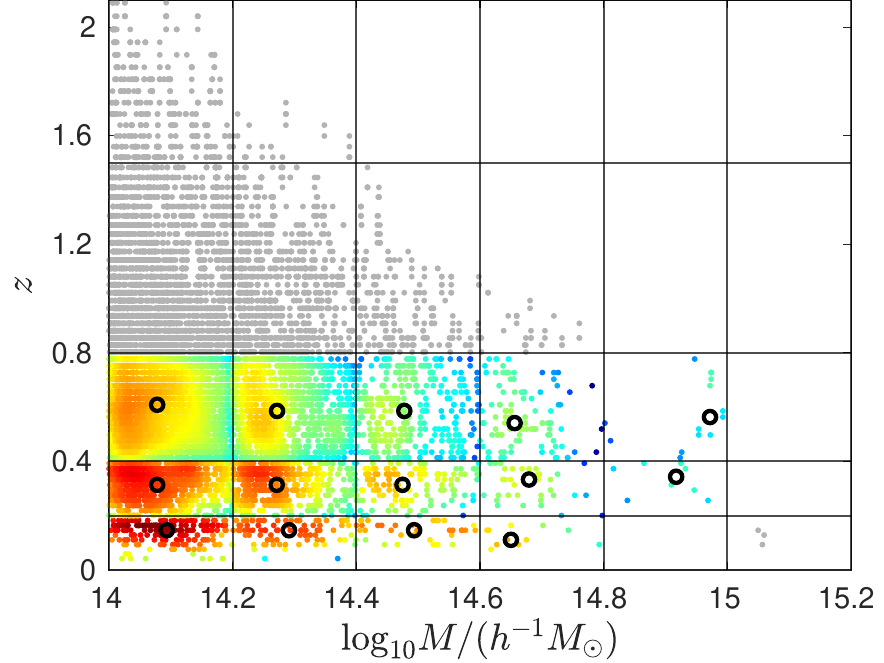}
    \caption{Bins of halos with ${\rm log}_{10}M>14$ (with $M$ in unit of $h^{-1}M_\odot$) in the mock {\it Euclid} data. 
Each dot presents a halo. Colors present the number density of halos in the bin. Black circles mark the median values of the mass and redshift as the representative values for
the corresponding bins. Grey dots are the halos not considered in our dilution+IA analysis.
    \label{fig:halobin}}
\end{figure}

To mimic observational analyses, here we follow the same strategy to measure the excess of satellites in our source sample around clusters.  Specifically, we identify all massive clusters  with $M\ge M_*$ in the mock simulation areas in accord with our model considerations with
high peaks being dominantly from massive clusters. We take $M_*=10^{14}h^{-1}M_{\odot}$. We then divide them into $5$ redshift bins from $0$ to $2.1$ and $6$ mass bins from ${\rm log}_{10}M/h^{-1}M_{\odot}=14$ to 15.2 as shown in Fig.\ref{fig:halobin}. 
Accounting for the lensing efficiency, clusters beyond $z=0.8$ contribute little to the high peaks, and we therefore do not consider their satellite IA effects. Also bins with very few massive clusters are not considered. 
The final bins used in measuring the satellite number boost are shown in colors in Fig.\ref{fig:halobin}, where the black circles indicate the median redshift and mass in different bins. 

\begin{figure}
    \plotone{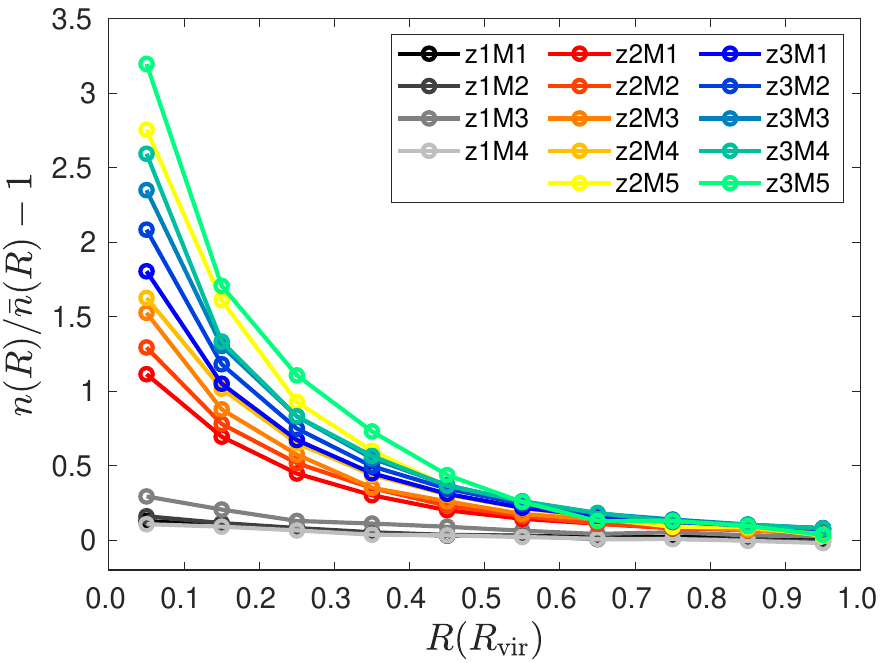}
    \caption{The mean excess galaxy number density profiles for each considered halo bin.
    \label{fig:excessfrac}}
\end{figure}

For each of these bins, based on their angular positions, we measure the 2D-excess of galaxies included in the source sample around each cluster, and take the  mean excess as the representative one of the considered bin. The excess profile is shown in Fig.\ref{fig:excessfrac}, where different colors are for different bins as shown in the legend. For example, z1M1 is for the lowest redshift and mass bin. 
The specific information of different bins is listed in Table.\ref{tab:halobin}, where the mean excess fraction is calculated by the ratio of the average number density of the satellite excess within a cluster region to the average number density of background galaxies.

\begin{table*}
    \centering
    \caption{Redshift and mass bins for measuring the satellite boost}
    \label{tab:halobin}
    \begin{tabular}{c c c c c c c}
    \toprule
    \multirow{2}{*}{bin label} & \multirow{2}{*}{redshift range} & mass range & \multirow{2}{*}{median $z$} & median mass & mean excess fraction & \multirow{2}{*}{Num of clusters}\\
    & & ($10^x M_\odot\,h^{-1}$) & &($10^{14} M_\odot\,h^{-1}$)& {\bf ($n(\leq R)/\bar{n}(\leq R)$)} & \\
    \hline
    {z1M1}& \multirow{4}{*}{$0<z\le0.2$}& $14.0<x\le14.2$ & 0.1460 & 1.2403 & 1.1331 & 189 \\
    {z1M2}& & $14.2<x\le14.4$ & 0.1464 & 1.9551 & 1.1161 & 90 \\
    {z1M3}& & $14.4<x\le14.6$ & 0.1464 & 3.1151 & 1.1421 & 35 \\
    {z1M4}& & $14.6<x\le14.8$ & 0.1111 & 4.4671 & 1.0871 & 11 \\
    \hline
    {z2M1}& \multirow{5}{*}{$0.2<z\le0.4$}& $14.0<x\le14.2$ & 0.3138 & 1.1967 & 1.2405 & 933 \\
    {z2M2}& & $14.2<x\le14.4$ & 0.3138 & 1.8673 & 1.2756 & 454 \\
    {z2M3}& & $14.4<x\le14.6$ & 0.3138 & 2.9833 & 1.2985 & 142 \\
    {z2M4}& & $14.6<x\le14.8$ & 0.3333 & 4.7778 & 1.3629 & 35 \\
    {z2M5}& & $14.8<x\le15.0$ & 0.3531 & 8.2643 & 1.4702 & 8 \\
    \hline
    {z3M1}& \multirow{5}{*}{$0.4<z\le0.8$}& $14.0<x\le14.2$ & 0.6088 & 1.1967 & 1.2465 & 3149 \\
    {z3M2}& & $14.2<x\le14.4$ & 0.5860 & 1.8699 & 1.2937 & 1244 \\
    {z3M3}& & $14.4<x\le14.6$ & 0.5860 & 3.0032 & 1.3426 & 385 \\
    {z3M4}& & $14.6<x\le14.8$ & 0.5413 & 4.5287 & 1.3831 & 105 \\
    {z3M5}& & $14.8<x\le15.0$ & 0.5635 & 9.3743 & 1.5415 & 15 \\
    \hline
    \end{tabular}
\end{table*}

\subsection{Single-halo simulation for satellite IA corrections}\label{sec:onehalo}

With the satellite excess information, we can analyze the change of the lensing term $K_{\rm{H}}$ due to the dilution and the IA effects, and their contributions to the 
noise field in cluster regions. For that, we perform single-halo
simulations with different dispersion $\sigma_{\theta}$ to obtain the correction factors of $\alpha_{\rm{dilution}}\alpha_{\rm{IA}}$ and $\sigma^2_{{\rm map},i} ({\rm halo})$.

This is done for each redshift and mass bin listed in Table.\ref{tab:halobin}, taking the median redshift and mass as the 
representative ones for setting up the halo simulation. For each single-halo simulation, 
we place the halo in the center of a $1.2\times1.2\ {\rm deg}^2$ map. Its 3D matter distribution follows the NFW profile
with mass-concentration relation from \cite{MC_Duffy08}. The source galaxies are first randomly populated over the map 
according to the source redshift distribution and the number density used in our mock simulations. We then re-distribute the
galaxies according to the corresponding galaxy excess information.

For that, we calculate the total excess number of satellite galaxies that should be in the halo, and pick this number of galaxies randomly from the region outside the halo virial radius. They are then placed inside the halo region following the excess profile shown in Fig.\ref{fig:excessfrac}.  

\begin{figure}
    \plotone{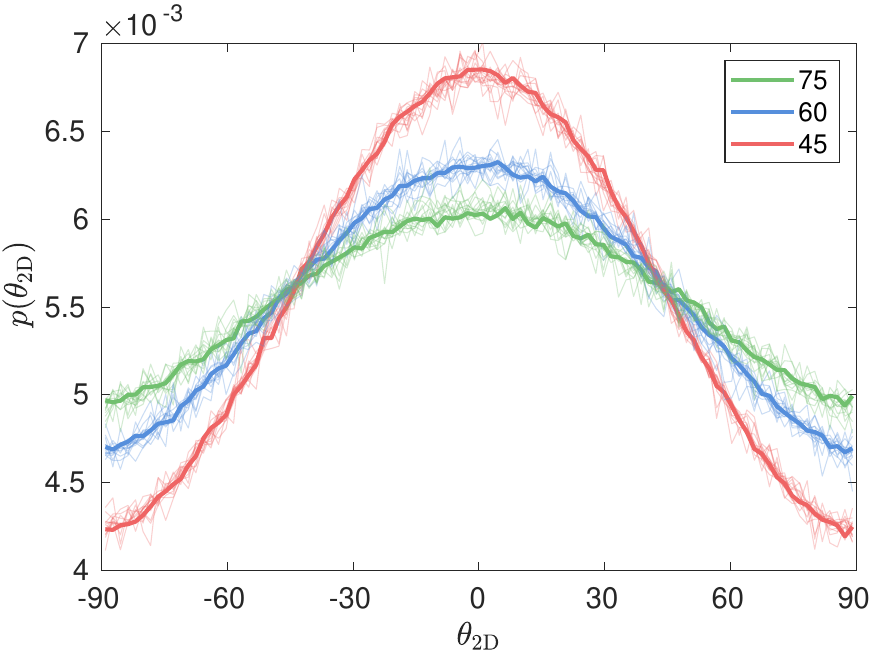}
    \caption{The projected 2D orientation distributions of satellites with respect to the cluster center from our shell calculations (thin) and from the mock simulations (thick) for the cases with $\sigma_\theta=75^\circ$ (green), $60^\circ$ (blue), and $45^\circ$ (red), respectively.
    \label{fig:ang_proj}}
\end{figure}

Because the excess is measured in the 2D space, we need the 2D IA information for the satellites. For our single-halo simulations, we consider the 3D IA model of Eq.\eqref{eq:theta_dis} with $\sigma_{\theta}$ being a constant in a cluster, which in a certain sense represents the average 3D IA. 
In this case, the 2D orientation distribution is the same for satellites populated in different 3D spherical shells around the cluster center, and thus independent of the 3D spatial distribution of the satellites. Therefore we can do the Monte Carlo calculations shell by shell. 
For each shell, we start from setting the 3D satellite orientations, and then perform the 2D projections to calculate their 2D orientation distribution. The specific procedures are the same as that used in
our mock simulations and are described in Appendix \ref{app:IAsetting}.
The results of the projection over all the spherical shells are shown in Fig.{\ref{fig:ang_proj}}, where the green, blue and red solid lines are for $\sigma_{\theta}=75^{\circ}$, $60^{\circ}$ and $45^{\circ}$, respectively. For each color,
different thin lines are for different considered mass and redshift bins. For comparison, we also show the corresponding distributions measured from our mock simulations in thick lines. They are consistent well.  As expected, with the increase of $\sigma_{\theta}$, $p(\theta_{\rm 2D})$ becomes wider. 
We smooth the distributions to obtain $p(\theta_{\rm 2D})$ for our 2D IA settings. 



With the 2D IA information, we can set satellites with IAs. We first assign intrinsic ellipticities to satellites using a Gaussian distribution with the total $\sigma_{\epsilon_{s}}=0.27$, in consistent with the mock simulations (see Sec.\ref{sec:mock}), and then rotate their orientations following the distributions given in Fig.{\ref{fig:ang_proj}}.
For each representative halo, we do $1000$ realizations of putting galaxies in the halo region and assigning IAs to them so that on average they are spherically distributed around the center of the halo.

To analyze the effects from satellites on WL signals, for each realization, we calculate the lensing signals $\kappa$ and $\bm{\gamma}$ from the halo on each background galaxy. 
The satellite galaxies have no lensing signals. We generate two sets of galaxy samples. Set 1 is for evaluate the average effect on $K_{\rm{H}}$. For that, we do not add intrinsic ellipticities 
to background galaxies, but the satellite galaxies do have intrinsic ones as described in the previous paragraph because we need to calculate their IA effects.  
Set 2 is for estimating the noise contributions from satellite galaxies with IAs, where we assign each background galaxy an intrinsic ellipticity following the Gaussian 
probability distribution with the total $\sigma_{\epsilon}=0.27$. 

\begin{figure}
    \gridline{\plotone{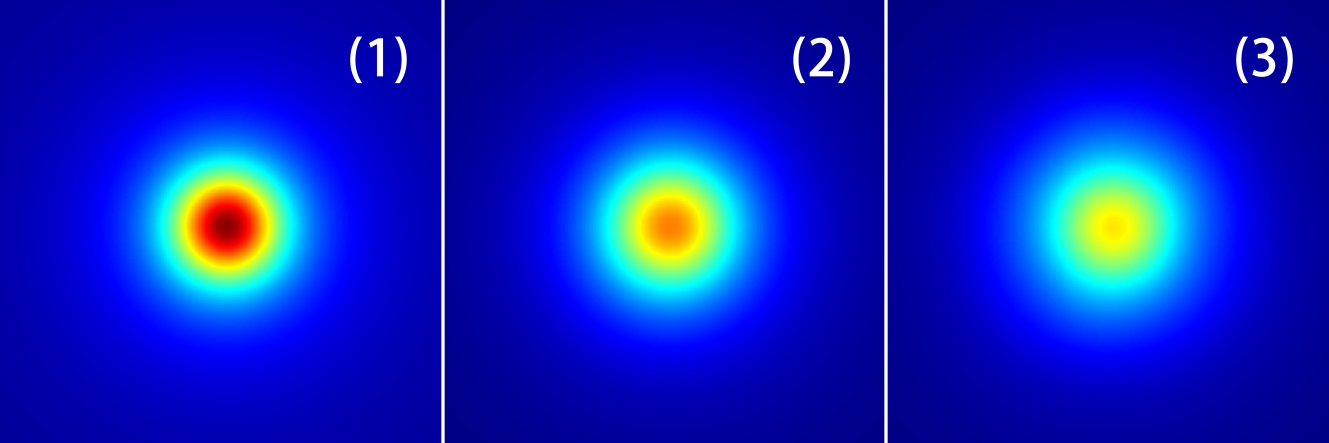}
    }
    \gridline{\plotone{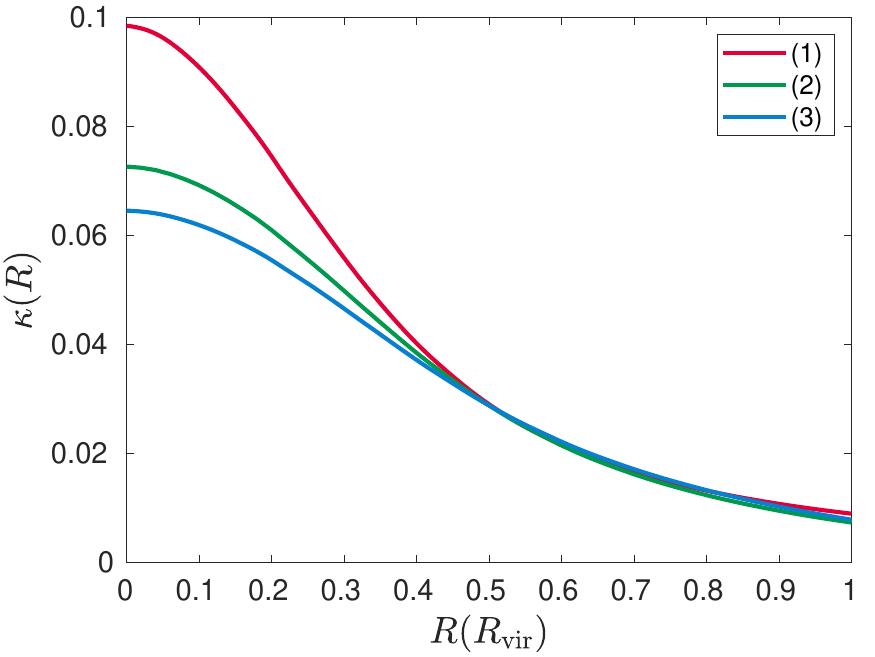}
    }
    \caption{Examples from Set 1 single-halo simulations with $\sigma_\theta=60^{\circ}$, and the bin z2M5. 
The upper panels show the central part of the reconstructed single-halo maps, and the lower panel plots the corresponding 1-D profiles of the halo regions.
Here (1) $K_{\rm NFW}$; (2) $K_{\rm NFW}\cdot\alpha_{\rm dilution}$; (3)$K_{\rm NFW}\cdot \alpha_{\rm dilution} \cdot \alpha_{\rm IA}$. 
    \label{fig:BIK}}
\end{figure}

For each realization of Set 1, we do the convergence reconstruction to obtain the lensing profile from the halo. We then take an average over the $1000$ realizations. 
We show an example in Fig.\ref{fig:BIK} for the bin of z2M5 and $\sigma_{\theta}=60^{\circ}$. The top panels show the average 2-D lensing profiles, where the left one is the
original $K_{\rm{H}}^{\rm{NFW}}$, the middle panel is for the profile including the dilution effect from the satellite excess but without the IA effect, i.e., $K_{\rm{H}}^{\rm{NFW}}\alpha_{\rm{dilution}}$,
and the right is for $K_{\rm{H}}^{\rm{NFW}}\alpha_{\rm{dilution}}\alpha_{\rm{IA}}$ with the IA effect included. The bottom panel plots the corresponding 1-D profiles.
We can see that both the dilution and the IA effects suppress the lensing signals in the central regions, 
leading to a lower peak. From Set 1 analyses, we obtain the profile of $\alpha_{\rm{dilution}}\alpha_{\rm{IA}}$ for each considered redshift and mass bin, which will be included in
correcting the model calculations. 

\begin{figure}
    \plotone{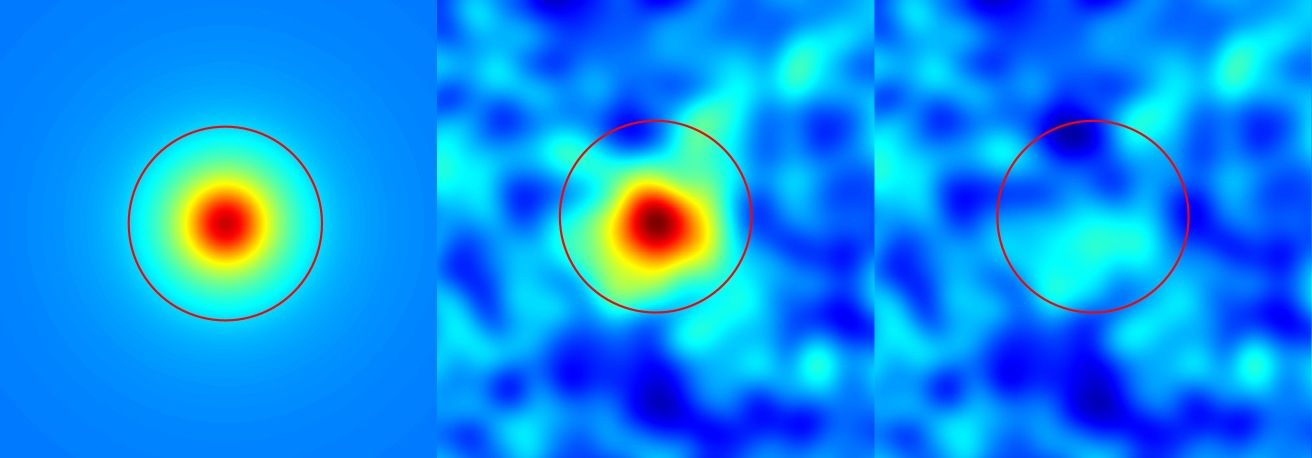}
    \caption{The examples from Set 2 single-halo simulations corresponding to that in Fig.\ref{fig:BIK}. The left panel is the noiseless case that is the same as the upper right panel of Fig.\ref{fig:BIK}. The middle panel is the noisy $K_N$ from reconstruction, and the right panel is the noise field $N_{\rm map}$. Red circles show the boundary of the halo region $R_{\rm vir}$. 
    \label{fig:KNnN}}
\end{figure}

To estimate the effects on the noise field in the halo region, we use Set 2 samples, and also perform the convergence reconstruction for each realization. An example is 
shown in Fig.\ref{fig:KNnN}. The left panel is the average lensing profile of the halo from Set 1 simulations that is the same as the top right panel of Fig.\ref{fig:BIK}.
The middle panel is from one realization of Set 2. The right one is the residual noise field $N_{\rm map}$ obtained by subtracting the left panel from the middle one.  
The red circle indicates the virial region of the halo. Within the halo, the noise contains the contributions from the random galaxy ellipticities and their correlation due to the satellite IAs.
The satellite number excess also takes an effect on the noise. From the residual maps of the $1000$ realizations for each bin, we calculate the noise parameters 
$\sigma^2_{{\rm map},i} ({\rm halo})$ for later model calculations.

\subsection{Theoretical model predictions and comparison with mock simulations}\label{sec:comparison}

From the single-halo simulations presented in Sec.{\ref{sec:onehalo}}, we derive the information of $\alpha_{\rm{dilution}}\alpha_{\rm{IA}}$ and 
$\sigma^2_{{\rm map},i} ({\rm halo})$ for different redshift and mass bins. They are functions of the satellite IA dispersion 
parameter $\sigma_{\theta}$. In Appendix \ref{app:1haloexamples}, we show the peak statistics from single-halo simulations with different mass and redshift taking $\sigma_{\theta}=60^{\circ}$ in Fig.\ref{fig:onehalopeak}, each with $1000$ realizations. 
The red and blue symbols are for the cases with and without dilution and IA effects, respectively. The lines are the corresponding model predictions. Here we have only one halo
with the known mass and redshift in the center plus the surrounding field region. By setting, there is no large-scale matter projection effect and no large-scale IAs.
We can see that our model works very well in the single-halo case.

For the full model calculations with IA effects, we also calculate their large-scale contribution to the noise parameters
$\sigma^2_{\rm{IA,i}}({\rm{NLA}})$ with Eqs.\eqref{eq:II}, \eqref{eq:GI} and \eqref{eq:Fz}. For the amplitude parameter $A_{\rm{IA}}$, we fit the large-scale simulation data with only
galaxy intrinsic ellipticities with IAs without lensing signals, i.e., the large-scale II correlations with the angular separation $\theta>5\hbox{ arcmin}$, 
and obtain the best-fit value of $A_{\rm IA}=0.8581$. This is used in the model calculations
in order to compare the theoretical predictions with the mock simulation data. As shown in \cite{Zhang2022}, the large-scale II is nearly independent of small-scale satellite IAs, thus 
this value is used for the model predictions with different $\sigma_{\theta}$.

We now have all the IA information ready for the full model calculations. We use the same cosmological parameters as the ones used in the simulation with 
$\Omega_{\rm m}=0.282$, $\Omega_{\Lambda}=0.718$, $\sigma_8=0.82$ and $h=0.697$. The halo mass function is taken from \cite{ShethTormen}, which is consistent with the halo mass function of our simulation that is based on ELUCID simulations \citep{Wang2016}.   
In using Eq.\eqref{eq:npeak_halo} to calculate the number of peaks in massive halo regions, we integrate over the halo mass and redshift. When they are in the range of
a redshift and mass bin listed in Table.\ref{tab:halobin}, the IA information ($\alpha_{\rm{dilution}}\alpha_{\rm{IA}}$ and $\sigma^2_{{\rm map},i} ({\rm halo})$)
of that bin is used in Eq.\eqref{eq:hatnpeak_halo}.

In real observations, the noise fields are obtained by randomly rotating galaxies that eliminate both the lensing 
signals and the IA signals. The noise parameters are then calculated by averaging over all the survey areas.  
This gives rise to only the random shape noise part $\sigma^2_{{\rm N},i}(\rm{average})$ from the average number density of source galaxies 
over the survey areas. The signal-to-noise ratio of peaks is then defined by using this $\sigma_{\rm{N},0}(\rm{average})$. 
In our mock simulation analyses, we follow the same procedures for peak statistics.  
In order to compare with mock data, in our model calculations, we therefore need to scale the signal-to-noise ratio of peaks in the cluster and field
regions with the factor $\sigma_{\rm{N,0}}(\rm{average})/\sigma_{\rm{total,0}}(\rm{halo})$ and
$\sigma_{\rm{N,0}}(\rm{average})/\sigma_{\rm{total,0}}(\rm{field})$, respectively, so that the predicted number of peaks is in accord 
with the \textquotesingle observed\textquotesingle\hbox{ }  signal-to-noise ratios.

\begin{figure}
    \plotone{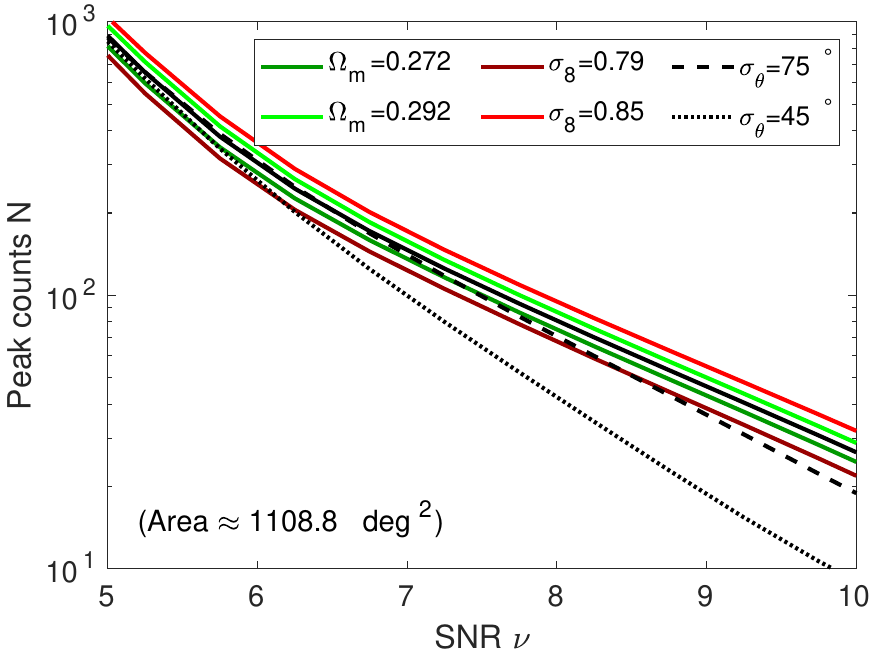}
    \caption{The cosmology and IA dependence of our theoretical peak model. The black solid, dashed and dotted lines are, respectively, calculated under the fiducial cosmology of our simulation including the dilution effect without IAs, with IAs of $\sigma_{\theta}=75^{\circ}$ and $45^{\circ}$. The colored lines are for cosmological models
    of varying $\Omega_{\rm m}$ or $\sigma_8$ as shown in the legend with the dilution effect included but without IAs.}
    \label{fig:peak_params}
\end{figure}

In Fig.\ref{fig:peak_params}, we show the model predictions for different $\Omega_{\rm m}$, $\sigma_8$ and $\sigma_{\theta}$. The black solid, dashed and dotted lines are for the fiducial cosmological model with the dilution effect but without IAs, with IAs taking $\sigma_{\theta}=75^{\circ}$, and $45^{\circ}$, respectively. The colored lines
correspond to the black solid line but with different $\Omega_{\rm m}$ or $\sigma_8$ as indicated in the legend with the other cosmological parameters fixed to the fiducial values. Here we emphasize again that our model is valid for high peaks that are dominantly originated from massive halos. For the considered Euclid-llike case, this corresponds to $\nu\ge \sim 5$ \citep[see also][]{FLAMINGOpeak}.

\begin{figure}
    \plotone{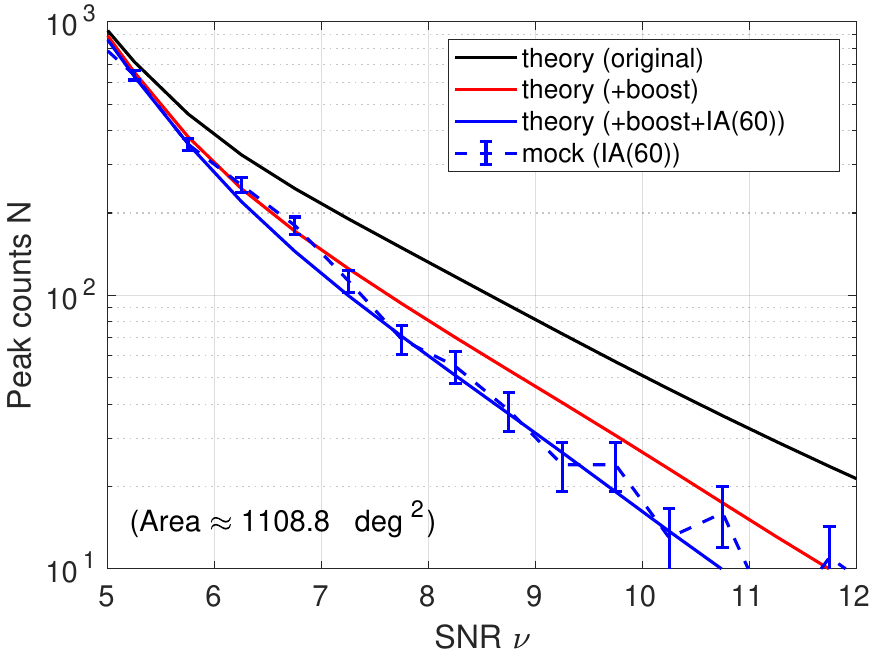}
    \caption{The comparison between the model predictions and the mock peak data in the case of $\sigma_\theta=60^\circ$. The black, red and blue lines are the model results
without considering the dilution+IAs effects, considering only the dilution effect without IAs, and considering the full corrections of dilution+IAs, respectively.
    \label{fig:theorycompare}}
\end{figure}

In Fig.\ref{fig:theorycompare}, we show our model predictions in comparison with the mock simulation data in the case of $\sigma_{\theta}=60^{\circ}$.
The black and red lines are the model results without the dilution and the IA corrections and with the dilution corrections only, respectively. 
The blue line is the predictions with the full corrections. 
It is seen that our fully corrected model predictions agree with the mock data very well where the error bars are the Poisson errors. Both the black and red lines deviate from the data significantly under the mock survey conditions, which
will lead to considerable biases if they are used to derive cosmological parameter constraints. To be shown later, with our corrected model, we are able to mitigate the bias and also extract the
information on $\sigma_{\theta}$ simultaneously from the \textquotesingle observed\textquotesingle\hbox{ } high peak counts. 

\begin{figure}
    \plotone{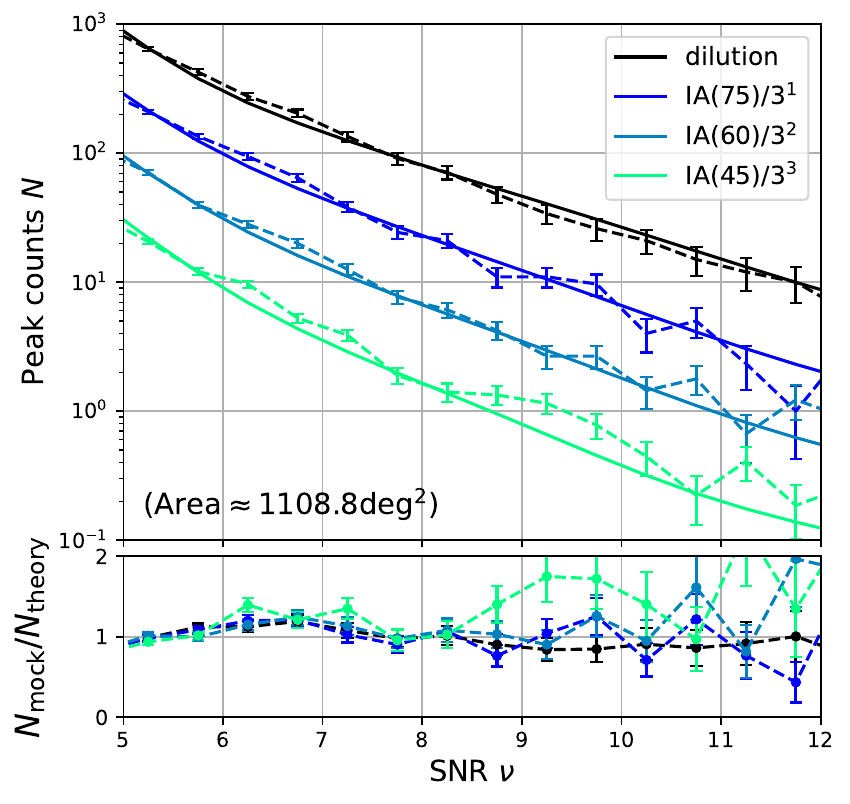}
    \caption{Comparisons of the model predictions (lines) with the mock data (symbols with error bars) in the cases of dilution only without satellite IAs (black), 
$\sigma_{\theta}=75^{\circ}$ (blue), $\sigma_{\theta}=60^{\circ}$ (dark green), and $\sigma_{\theta}=45^{\circ}$ (light green), respectively. 
For clarity, the peak counts are scaled as shown in the legend in the upper panel. The lower panel shows the relative differences of the mock data to the model predictions.
    \label{fig:peakcounts}}
\end{figure}

In Fig.\ref{fig:peakcounts}, we further compare our model predictions with the mock data in cases of $\sigma_{\theta}=75^{\circ}$ (blue), $60^{\circ}$ (dark green) and $45^{\circ}$ (light green), respectively.
For clarity, in the upper panel, we scale the counts in different cases as shown in the legend. The lower panel shows the relative differences between the mock data and the model predictions.
The black line (model) and the data with error bars are for the case without satellite IAs but only the effects due to the satellite galaxy number boost ($\alpha_{\rm{dilution}}$ + noise change in the model calculations). 
We can see that in all the cases, our model gives rises to predictions that are consistent well with the mock data. In the case of $\sigma_{\theta}=45^{\circ}$, the mock data are somewhat higher than the model results for SNR $\nu>8$.
We notice that a significant fraction of these peaks correspond to the very high peaks in the dilution-only case (black). Due to the strong satellite IAs with $\sigma_{\theta}=45^{\circ}$, they are suppressed and shift to lower SNRs. These peaks
can be affected by the sample variance of our simulations which are constructed from a single N-body simulation with the size of $500h^{-1}\hbox{Mpc}$, leading to relatively large deviations from the model predictions. 
We also note that there are observational indications that the satellite IAs may not be very strong\citep{Sifon,DESY1redmapperIA}, at the level corresponding to $\sigma_{\theta}\sim 70^{\circ}$ \citep{Knebe2008, SB2010, Zhang2022}. 
In that range, our peak model works very well.



\subsection{The bias mitigation on the cosmological constraints with our IA-corrected model}\label{sec:constraint}

To further validate our model, here we show how it can be used to derive cosmological constraints and to mitigate the IA-induced bias from high peak counts.
We consider two sets mock data with $\sigma_{\theta}=75^{\circ}$ and $60^{\circ}$, respectively. For each data set, we perform two Markov Chain Monte Carlo (MCMC) constraints, one using the model
without the dilution and IA corrections, and the other one with our fully corrected model. To show the ability to control the IA-induced bias, here we assume that the underlying $\sigma_{\theta}$ is known in our 
IA-corrected model calculations and perform constraints only on the cosmological parameters $\Omega_{\rm m}$ and $\sigma_8$. In the next subsection, we will forecast the three-parameter constraints
on $\Omega_{\rm m}$, $\sigma_8$ and $\sigma_{\theta}$ using Fisher calculations, and also discuss how to make our model a theoretical tool in future MCMC constraints.  

It is noted that the analyses of our single-halo simulations assume the fiducial cosmological model. We apply the derived $\alpha_{\rm{dilution}}\alpha_{\rm{IA}}$ to
all the cosmological models in the MCMC calculations. In principle, $\alpha_{\rm{dilution}}\alpha_{\rm{IA}}$ is cosmology dependent through the virial radius of 
a massive cluster for estimating the satellite boost, and through the density profile of the cluster. Because the satellite boost in the inner region of the cluster  
affects the cluster lensing signal the most and the cosmology dependence of the M-c relation of dark matter halos is relatively weak, the simplified treatment adopted here
should not have significant impacts on our results. We leave the detailed analyses on the cosmology dependence of $\alpha_{\rm{dilution}}\alpha_{\rm{IA}}$ in our future studies. 
On the other hand, we take into account the dependence of the large-scale II and GI terms on cosmological parameters in our MCMC fitting.

We derive the  MCMC constraints by minimizing the following $\chi^2$,
\begin{equation}
    \label{eq:chisqr}
    \chi^2=\sum_{i,j}^{N_{\rm bin}}\Delta N_i \widehat{C_{ij}^{-1}} \Delta N_j,
\end{equation}
where $\Delta \bar{N}_i=N_i^{(t)}-N_i^{(d)}$ is the difference of the peak number from theoretical prediction under a trail cosmology and the peak number from data (the mock observational data in this case) in the $i$-th SNR $\nu$ bin. The covariance matrix $C_{ij}$ is estimated by adopting the bootstrap method as
\begin{equation}
    \label{eq:Cov}
    C_{ij}=\frac{1}{R-1}\sum_{r=1}^{R}(N^{(d,r)}_i-\bar{N}^{(d)}_i)(N^{(d,r)}_j-\bar{N}^{(d)}_j),
\end{equation}
where $\bar{N}^{(d)}$ is the mean of the total $R=1000$ bootstrap samples, and $N^{(d,r)}$ is the $r$-th bootstrap sample. We note that
we do the bootstrap using the $156$ patches of the maps each with $3\times 3\hbox{ deg}^2$. Thus our estimation of the covariance does not include large-scale correlations beyond
the map size. This may underestimate the covariance, but not significantly for the high peaks concerned in our study here.  
We follow \citet{Hartlap2007} to calculate the unbiased inverse of the covariance matrix,
\begin{equation}
    \label{eq:Covinv}
    \widehat{C_{ij}^{-1}}=\frac{R-N_{\rm bin}-2}{R-1}C_{ij}^{-1}.
\end{equation}

\begin{table}
    \centering
    \caption{Marginalized MCMC constraints on the cosmological parameters from the mock data using our theoretical peak models without and with dilution+IA corrections.}
    \label{tab:MCMC}
    \begin{tabular}{c c c c c}
    \toprule
     & $\sigma_\theta$ & $\Omega_{\rm m}$ & $\sigma_8$ & $S_8$ \\
    \hline
    fiducial & - & 0.282 & 0.82 & 0.795\\
    \hline
    \multirow{2}{*}{IAin} & 75 & $0.287^{+0.044}_{-0.04}$ & $0.805^{+0.063}_{-0.054}$ & $0.8^{+0.0048}_{-0.0054}$ \\
     & 60 & $0.231^{\pm 0.048}$ & $0.898^{+0.098}_{-0.085}$ & $0.801^{+0.0058}_{-0.0065}$ \\
    \hline
    \multirow{2}{*}{IAno} & 75 & $0.369^{+0.041}_{-0.044}$ & $0.681^{+0.044}_{-0.035}$ & $0.76^{+0.0042}_{-0.0048}$ \\
     & 60 & $0.401^{+0.037}_{-0.04}$ & $0.648^{+0.035}_{-0.029}$ & $0.753^{+0.0044}_{-0.0049}$ \\
    \hline
    \end{tabular}
\end{table}
\begin{figure}
    \plotone{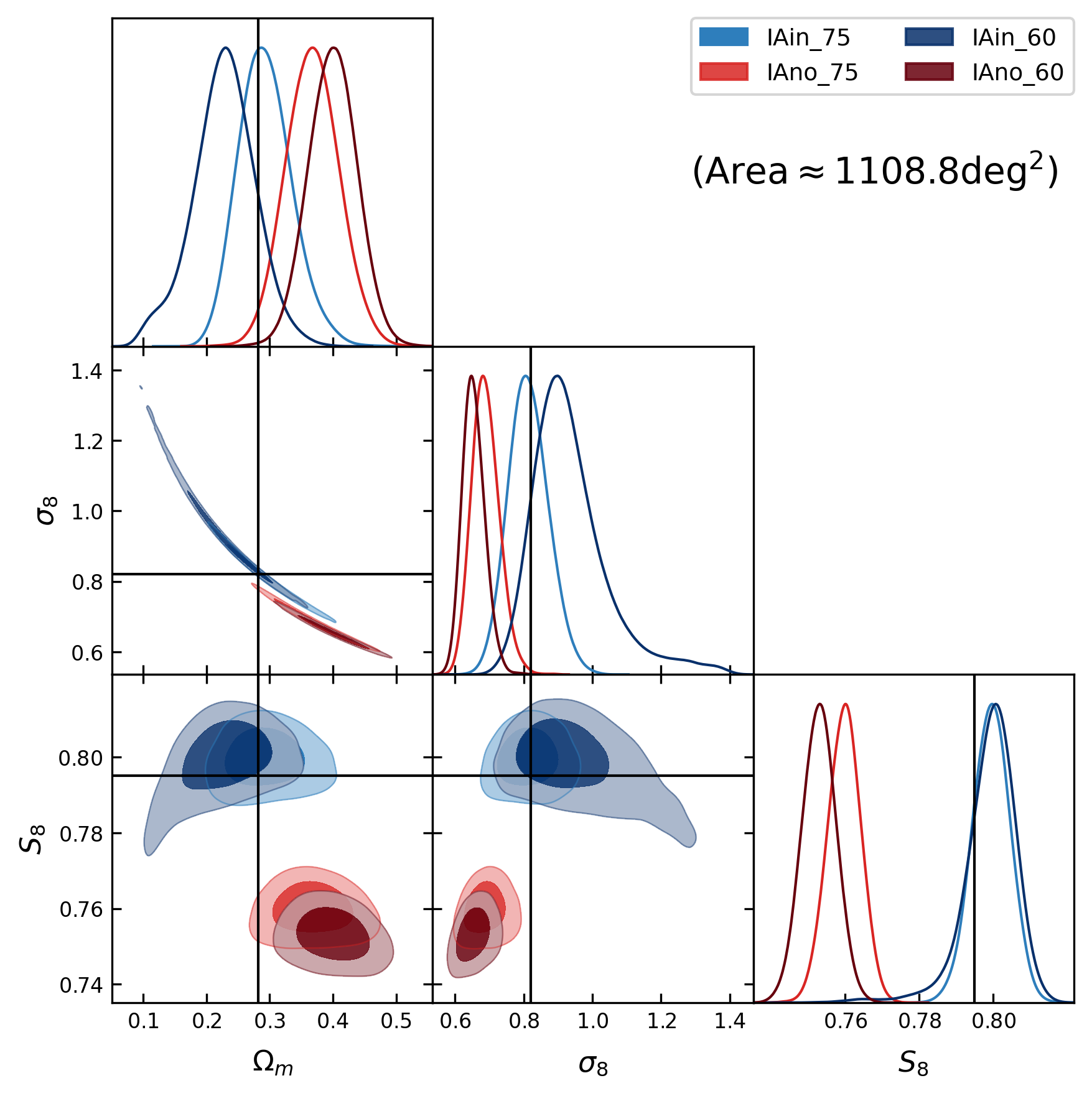}
    \caption{The MCMC constraints from the mock data using the models without (red series) and with (blue series) dilution+IA corrections.
    \label{fig:MCMC}}
\end{figure}

For the mock peak data, we consider peaks with the SNR in the range of $[5,11.5]$ and $[5,9.5]$ for the cases of $\sigma_{\theta}=75^{\circ}$ and $60^{\circ}$, respectively. The lower cut corresponds to
the validity of our model for high peaks.  The cut on the high SNR end is to avoid the fluctuations due to the sample variance of the simulations.

We consider $(\Omega_{\rm m}, \sigma_8)$, and fix all the other cosmological parameters to the values used in the simulations. Flat priors of $([0.05,0.95],[0.20,1.60])$ for $(\Omega_{\rm m},\sigma_8)$ are applied 
in the MCMC constraints. The results are shown in Fig.\ref{fig:MCMC}, where the red and the dark red ones are for the cases of $\sigma_{\theta}=75^{\circ}$ and $60^{\circ}$, respectively, using the peak model 
without including the dilution and the IA effects, and the blue and the dark blue ones are the corresponding constraints using the fully corrected peak model.  
We can see that without considering the corrections, the derived $S_8=\sigma_8(\Omega_{\rm m}/0.3)^{0.5}$ parameter is significantly biased to lower values. 
As shown in Table.\ref{tab:MCMC}, in the mock survey condition with the Euclid/CSST-like source redshift distribution and the survey area of $\sim 1000\hbox{ deg}^2$, the bias can reach $(0.795-0.76)/0.0042\sim 8\sigma$
for $\sigma_{\theta}=75^{\circ}$. For $\sigma_{\theta}=60^{\circ}$, the bias is $\sim 10\sigma$. 
With our fully corrected model, the $S_8$ constraints are $0.80^{+0.0048}_{-0.0054}$ and $0.801^{+0.0058}_{-0.0065}$, respectively, for $\sigma_{\theta}=75^{\circ}$ and $60^{\circ}$, which are within $1\sigma$ range of the fiducial $S_8$ value.
The results demonstrate well the important potential of our model to control the IA-induced biases on the cosmological parameter constraints.

\subsection{Potential constraint on the IA parameter $\sigma_{\theta}$}\label{sec:Fisher}

With our IA corrected model, it is also possible to constrain the satellite IA parameter $\sigma_{\theta}$ simultaneously with the cosmological parameters from the observed high peak counts.
To show the potential, here we perform Fisher analyses using the Fisher matrix 
\begin{equation}
    \label{eq:Fisher}
    F_{\alpha\beta}=\sum_{ij}\frac{\partial N^{(t)}_i}{\partial p_\alpha} \widehat{C_{ij}^{-1}} \frac{\partial N^{(t)}_j}{\partial p_\beta},
\end{equation}
where $p_\alpha$ is one of the model parameters $(\Omega_{\rm m},\sigma_8,\sigma_\theta)$, and the derivatives are estimated numerically for each peak bins using our IA corrected model. 
We note that the derivatives with respect to $\Omega_m$ and $\sigma_8$ can be calculated accurately from our model because they are explicit in the model setting. The derivative with respect to $\sigma_{\theta}$, however, is calculated based on our single-halo simulations. 
We test the stability of the Fisher results using different intervals of $\Delta(\sigma_{\theta})$ to calculate the derivative. They are stable for $\Delta(\sigma_{\theta})=15^{\circ}, 10^{\circ}$ and $5^{\circ}$. When $\Delta(\sigma_{\theta})<5^{\circ}$, 
the derivative cannot be computed properly because the model differences for such small $\Delta(\sigma_{\theta})$ are small given the finite number of satellite excess in clusters. We thus take $\Delta(\sigma_{\theta})=5^{\circ}$ in our calculations.

\begin{figure}
    \plotone{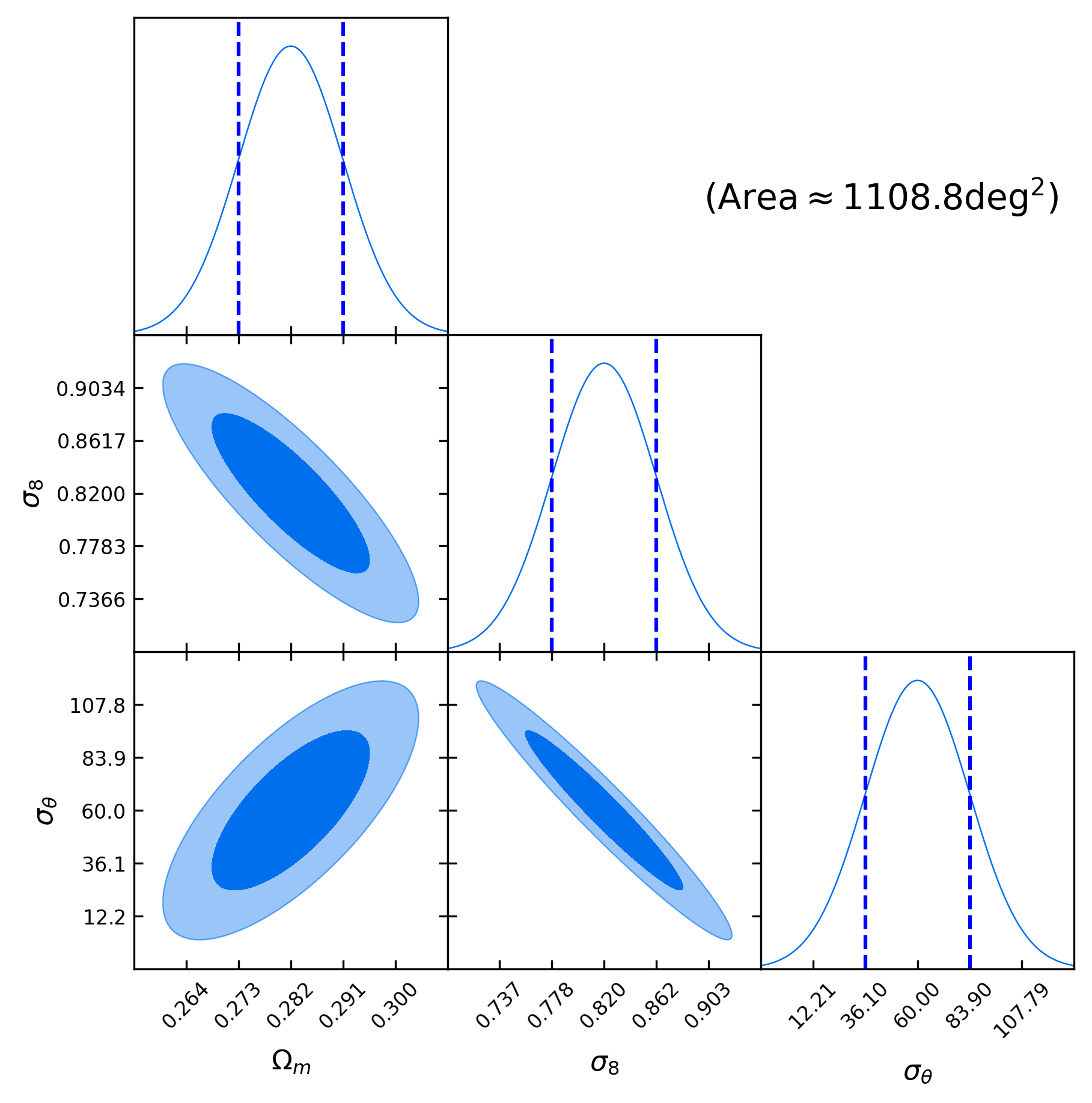}
    \caption{The Fisher forecast for the model parameters. The fiducial value is $\sigma_\theta=60^\circ$. Dashed lines indicate the $1\sigma$ of the 1-D distributions.
    \label{fig:Fisher}}
\end{figure}

As an example, the results for the case of $\sigma_\theta=60^\circ$ case are shown in Fig.\ref{fig:Fisher}. 
We see that there is a strong degeneracy between $\sigma_8$ and $\sigma_{\theta}$. As we show in previous sections, the high peak counts of ${\rm SNR}\ge 5$ we considered are affected by IA sensitively. With higher IA (smaller $\sigma_{\theta}$), the peak counts are systematically more suppressed. Thus given a set of high peak counts, 
the effect from smaller $\sigma_{\theta}$ needs to be compensated by higher $\sigma_8$, leading to the degeneracy seen in Fig.\ref{fig:Fisher}. The different degeneracy direction between $\Omega_m$ and $\sigma_{\theta}$ is attributed to the strong degeneracy between $\Omega_m$ and $\sigma_8$. With $\sigma_8$ fixed, 
smaller $\sigma_{\theta}$ also needs higher $\Omega_m$ to maintain the peak counts unchanged, and the degeneracy direction of $\Omega_m$ and $\sigma_{\theta}$ is opposite to the one shown in Fig.\ref{fig:Fisher}.
The 3-parameter Fisher results show that the $1\sigma$ constraint on $\sigma_{\theta}$ is 
$\sim 24^{\circ}$ with the considered survey conditions. By scaling the survey area from $\sim 1000\deg^2$ to $\sim 15000\deg^2$, the projected constraint can reach $\sigma(\sigma_{\theta})\sim 6^{\circ}$ for the satellite IA parameter $\sigma_{\theta}$.

In order to perform MCMC constraints with $\sigma_{\theta}$ also as a free parameter, we need to be able to predict the high peak counts given an arbitrary $\sigma_{\theta}$. Note that we obtain 
$\alpha_{\rm{dilution}}\alpha_{\rm{IA}}$ and $\sigma^2_{{\rm map},i} ({\rm halo})$ from single-halo simulations with the satellite boost information 
measured from observations. It is inefficient to perform such simulations
for each set of $(\Omega_{\rm m}, \sigma_8, \sigma_{\theta})$ in situ during the MCMC runs. A better approach is to do a number of single-halo simulations with different $\sigma_{\theta}$ 
that are relatively densely sampled in the given prior range. We then can theoretically calculate the peak numbers in a parameter grid of $(\Omega_{\rm m}, \sigma_8, \sigma_{\theta})$. From that, 
we can do interpolations to build an emulator to predict the peak numbers for any given $(\Omega_{\rm m}, \sigma_8, \sigma_{\theta})$ for MCMC constraints. 
We will carry out such analyses in our forthcoming studies.

\subsection{Potential systematic effects}\label{systematics}

In our analyses, we use the mock simulations to mimic real observations to demonstrate the applicability of our theoretical model. However, there are observational effects not included in the simulations, notably the center offset of central galaxies and 
the uncertainties in cluster mass determinations. With cluster redshifts estimated photometrically (photo-z), uncertainties also exist. These can affect our estimations of the satellite excess in massive clusters.  Considering photo-zs, their estimates for massive clusters
typically have higher accuracies reaching $\sigma_z/(1+z) \sim 0.01$ comparing to $\sim 0.05$ for individual field galaxies in the current generation of surveys \citep[e.g.,][]{Rykoff2016}. 
In our analyses, the lowest redshift bin width used to measure the satellite excess is $0.2$, much wider than the photo-z uncertainties. We therefore expect insignificant impacts from photo-z uncertainties. For the center offset and 
the cluster mass uncertainties, here we perform additional analyses to investigate their effects on our studies. 

\begin{figure}
    \plotone{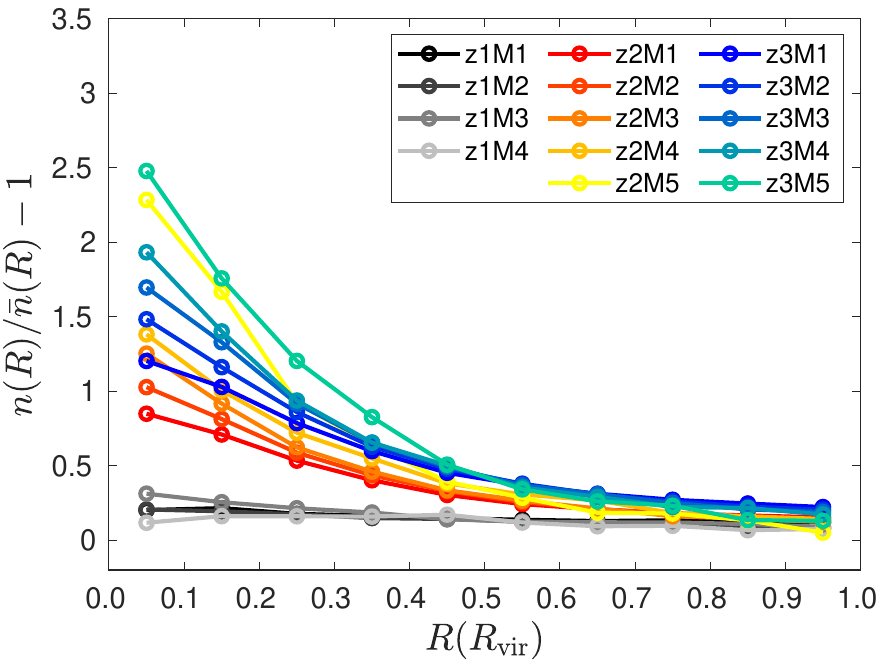}
    \caption{Same as Fig.\ref{fig:excessfrac} but taking into account the off-centering effect.
    \label{fig:excessfrac_offset}}
\end{figure}

Regarding the center offset, we randomly pick an offset position for each considered cluster around which we measure the satellite excess. The offset distribution is chosen as follows 

\begin{equation}
        p(r)=f_{\rm cen}\frac{r}{\sigma_{\rm w}^2}\exp\left(-\frac{r^2}{2\sigma_{\rm w}^2}\right)
        +(1-f_{\rm cen})\frac{r}{\sigma_{\rm o}^2}\exp\left(-\frac{r^2}{2\sigma_{\rm o}^2}\right),
\end{equation}
where $f_{\rm cen}=0.92$ is the well-centered fraction, $\sigma_{\rm w}=90\ h^{-1}{\rm kpc}$ and $\sigma_{\rm o}=420\ h^{-1}{\rm kpc}$ are adopted for the well-centered (w) and the large offset (o) cases, respectively, in accord with the optically selected central galaxy offset distribution
 \citep{Rykoff2016,Oguri2018, Ding2025}. The excess profile in each considered cluster bin including center offset is shown in Fig.\ref{fig:excessfrac_offset}. Comparing to Fig.\ref{fig:excessfrac}, the off-centering effect induces statistically a smoothing on the satellite excess. 
 
\begin{figure}
    \plotone{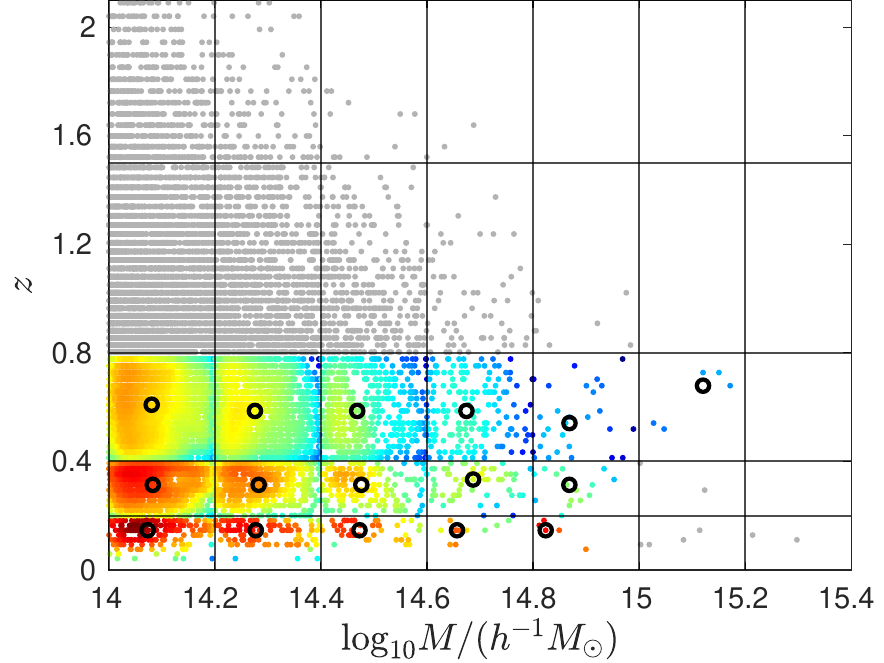}
    \caption{Same as Fig.\ref{fig:halobin} but with the uncertainties of mass determinations included.
    \label{fig:halobin_Merror}}
\end{figure}

\begin{figure}
     \plotone{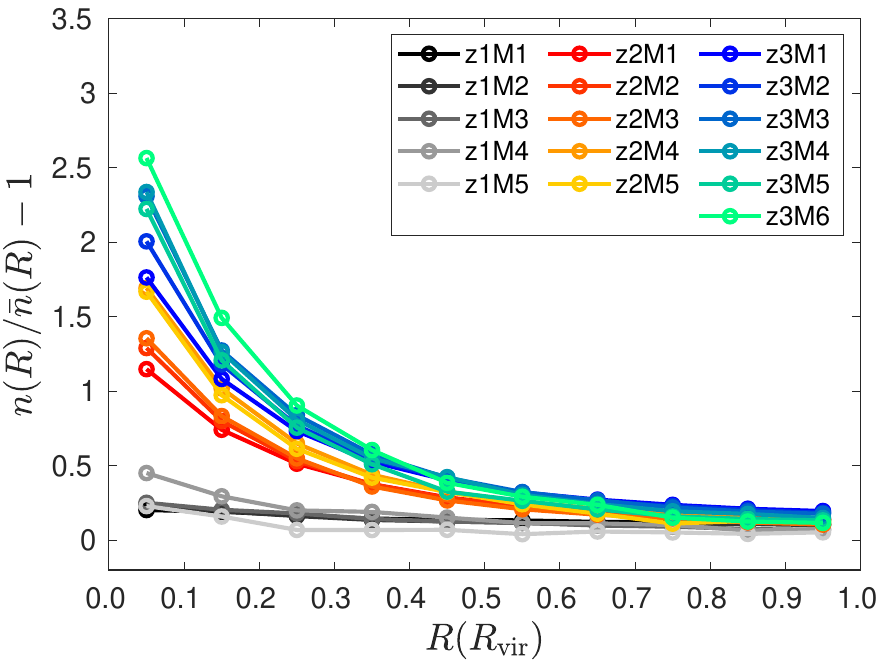}
    \caption{Same as Fig.\ref{fig:excessfrac} but including the mass uncertainties of clusters.
    \label{fig:excessfrac_mass}}
\end{figure}

 For the uncertainties of cluster mass determination, we adopt the distribution below from the richness estimates \citep[e.g,][]{Rykoff2012, WZL2015}
 
 \begin{equation}
    P(M)=\frac{1}{\sqrt{2\pi}\sigma_{\ln M}}\exp\left[-\frac{(\ln M-\ln M_{\rm tr})^2}{2\sigma^2_{\ln M}}\right],
\label{eq:massdis}
\end{equation}
where $M_{\rm tr}$ is the true mass of a clusters and $\sigma_{\ln M}=0.33$ \citep{Rykoff2012}. We first randomly perturb the mass of a cluster based on Eq.\eqref{eq:massdis}. Then we divide clusters into different bins based on their perturbed mass, as shown in Fig.\ref{fig:halobin_Merror}. 
Statistically, because the less massive clusters are more abundant, the mass uncertainties lead to more clusters outside the left side a bin into the bin than shifting clusters outside the right side of the bin. Thus we have more clusters with $M\ge 10^{14}h^{-1}M_\odot$ here resulting two more bins 
used in estimating the satellite excess comparing with Fig.\ref{fig:halobin}. The corresponding excess profiles are shown in Fig.\ref{fig:excessfrac_mass}.


\begin{figure}
    \plotone{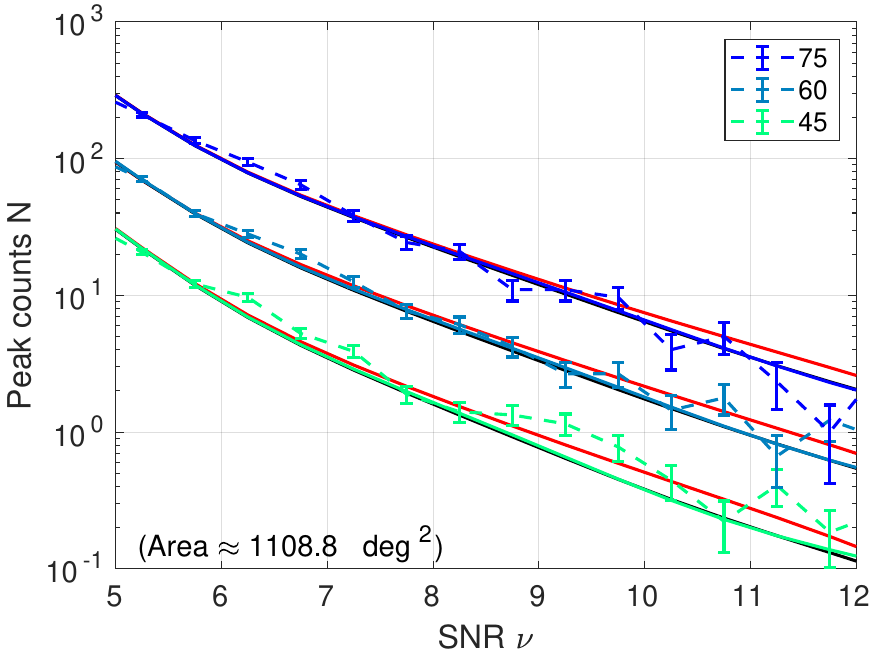}
    \caption{Comparison of the model predictions without the center offsets and without the mass uncertainties (blue, dark green and green), with the center offsets (black) and with the mass uncertainties (red), respectively. The data points with error bars are from our mock simulations.
    \label{fig:peak_systematics}}
\end{figure}

With these 2D excess profiles, we repeat the single-halo simulations to obtain the information of $\alpha_{\rm{dilution}}\alpha_{\rm{IA}}$ and $\sigma^2_{{\rm map},i} ({\rm halo})$ for different mass and redshift bins for a given $\sigma_{\theta}$, and calculate the model predictions
for the high peak counts. We show the results in Fig.\ref{fig:peak_systematics} together with the data from the mock simulations. The blue, dark green and green lines are the model predictions without including the off-centering and the mass uncertainties (the same as those in Fig.\ref{fig:peakcounts} ), 
the black and red lines are the model predictions with the off-centering effect and the mass uncertainties, respectively. 

It is seen that the effect from the center offset is minimum here because we adopt a smoothing scale $\theta_G=2\hbox{ arcmin}$ in our analyses, which is in general larger than the off-centering induced smoothing in the excess estimations. For the mass uncertainties, they can affect 
model predictions, especially in the very high peak part. For the survey conditions considered here, the differences are still within the data error bars. However, for larger surveys with smaller statistical errors, we need to consider the cluster mass uncertainties more carefully. On the other hand, we also expect 
that the mass determination from different cluster proxies can be improved in the future. 

We note that in measuring the satellite excess in our mock simulations, we use the lensing-shifted positions of galaxies \citep{Wei}, and thus the lensing magnification effect is partially taken into account. However, we select source galaxies only based on their redshift distributions without 
considering their magnitude cut, and therefore the luminosity-related magnification effect is not accounted for here. We will investigate the full magnification effect in our future studies. 

It is also noted that here for the single massive clusters, our model only consider their host NFW halos in calculating their
lensing signals without including the baryonic effects. The model predictions agree with our mock simulations well because the simulations model the galaxy formation semi-analytically with no baryonic feedbacks. On the other hand, within the model framework, we can make modifications to account for the baryonic effects. 
If they can be phenomenologically described as affecting the halo profile through the mass-concentration relation \citep{Mead2015}, our model can be readily extended to include such effects by allowing the mass-concentration relation being different from that of pure dark matter halos \citep[e.g.,][]{LiuX2015} and 
by modifying the large-scale projection effects with the baryonic effects included \citep[e.g.,][]{Mead2015}. In \cite{LZW}, we analyze the peak steepness statistics where the steepness is measured by the second derivatives of a peak. It is shown that this statistics is more sensitive to the halo profile than the peak height statistics. By combining 
the two statistics, we can constrain the amplitude of the mass-concentration relation simultaneously with the cosmological parameters, and thus potentially probe the baryonic feedbacks from weak lensing peak statistics (Li et al. in preparation). 

Furthermore, being one of our major efforts in the future, we will include the baryonic effects by adopting, for example, the Baryon Correction Model  \citep[BCM,][]{BCM2020} 
to describe the mass distribution of a cluster.  The halo mass function should also be modified correspondingly. With this, we can model the impacts of the baryonic effects on weak lensing high peak statistics more physically. 
It is pointed out that with these changes, our method presented here to model the IA impacts should still be valid as long as the proper density profile of massive clusters is adopted.

\section{Summary and discussions} \label{sec:discussion}\

In this paper, we present a theoretical model for WL high peak counts with the IA effects included. They come into the model from two parts. One is to modify the 
lensing profile of massive clusters of galaxies due to the satellite IAs. For that, the satellite number boost in the source sample around massive clusters needs to be measured 
from data, thus it is survey specific. With the boost information, the correction factor $\alpha_{\rm{dilution}}\alpha_{\rm{IA}}$ can be derived for a given alignment dispersion angle $\sigma_{\theta}$
from single-halo simulations. The other IA effect is to change the noise parameters. For that, we consider separately the massive cluster regions and the large scale part. In the massive cluster regions, 
the noise parameters $\sigma^2_{{\rm map},i} ({\rm halo})$ can also be obtained from single-halo simulations. The large scale IA contributions to the noise are calculated from the 
cosmological-dependent NLA model. 

We validate the model by comparing with mock simulation data. We also show how the model can control the IA-induced bias on cosmological parameter constraints. For Euclid/CSST-like source redshift distribution and 
the survey area about $1000\deg^2$, the $S_8$ bias is about $8\sigma$ even for relatively weak satellite IAs with $\sigma_{\theta}=75^{\circ}$ if the IA effects are not accounted for in the model calculations. 
With the IA effects included, our model can dramatically reduce the bias on $S_8$ to be less than $1\sigma$.  

Besides cosmological parameters, our model contains two IA parameters, the satellite IA strength represented by $\sigma_{\theta}$ and the amplitude parameter $A_{\rm{IA}}$ of the NLA model. 
This provides a theoretical framework to extract the IA parameters simultaneously with cosmological constraints from WL high peak statistics. 
In this paper, we use Fisher analyses to demonstrate the potential to constrain $(\Omega_{\rm m}, \sigma_8, \sigma_{\theta})$ using our model with $A_{\rm{IA}}$ fixed.  
We note that the dependence on $A_{\rm{IA}}$ is explicit in the model, thus it should be straightforward to include it as a free parameter in the fitting.
On the other hand, for the dependence on $\sigma_{\theta}$, single-halo simulations are needed with the boost information measured from data, and thus it is implicit.  
To make it explicit for MCMC constraints from observational data, we can build an emulator at the peak count level based on our model calculations. For that, we need to 
do single-halo simulations for relatively densely sampled $\sigma_{\theta}$ and compute corresponding peak counts. Then interpolation can be done to obtain the peak counts for any given $\sigma_{\theta}$.
It is more efficient to construct such a model-based emulator than that relying fully on large simulations.
This will be our future efforts. The cosmological dependence of $\alpha_{\rm{dilution}}\alpha_{\rm{IA}}$ will also be investigated, which can be taken into account in building the emulator. 


In our current study, we assume that the IAs are the same for all the satellite galaxies without considering the possible dependence on galaxy properties, such as the luminosity, type, anisotropic distribution, and the distance to the
center of the host cluster \citep[e.g,][]{Huang2018,Knebe2020,Tenneti2021, Fortuna2021, Lan2024, PAU2025, mTNGIA2025}. Given a good knowledge of these complications, they can be accounted for in our single-halo simulations to further improve
the model accuracy. We will work on this in future studies. 

For our general theoretical framework, the basic physical ingredients are the halo mass function, halo density profile and the large-scale projection effects. As shown in this study, the IA effects can be incorporated by modifying the lensing profiles of massive halos as well as the noise properties.  
The baryonic effects can also be modeled by changing these ingredients as discussed in the previous section.  
Besides these cosmological dependent quantities, there is also a nuisance parameter $M_*$ representing the mass limit of massive halos in our considerations. From physical considerations and different simulation studies, it should be $\sim 10^{14}h^{-1}M_\odot$.  
Being clear physically, however, the uncertainties of these ingredients, physical or nuisance, can affect our model predictions. In our previous studies, we investigate partially the sensitivities of our model predictions on these quantities. For example, in \cite{Yuan2018}, 
we show that choosing different values of $M_*$ around $\sim 10^{14}h^{-1}M_\odot$ can affect the model predictions somewhat mainly for the lowest side of the high peaks. The best value of $M_*$ may also be cosmological dependent. \cite{LZW} study the model dependence on the 
mass-concentration relation of massive halos, showing that the peak height statistics is less sensitive to it than the peak steepness statistics. In our forthcoming studies, we will investigate thoroughly the impacts of the uncertainties involved in our modeling. With that we can improve our model accuracies by 
including the uncertainties in the model calculations. We also aim to improve our model by calibrating the relevant parameters using simulations, similar to the approach adopted in calculating the nonlinear power spectrum with the halo model as the underlying considerations \citep[e.g.,][]{Takahashi2012, Mead2015, Mead2021}. 
Such improvements are important for applying weak lensing statistics to derive accurate cosmological information using data from future large surveys.

\begin{acknowledgements}
\modulolinenumbers[17]
This research is supported by  National Key R\&D Program of China No. 2022YFF0503403 and the NSFC grant No.11933002. Z.H.F. also acknowledges the support from NSFC grant No.U1931210, and the grant from the China Manned Space Projects with No. CMS-CSST-2021-A01 and CMS-CSST-2025-A05. X.K.L. acknowledges the support from NSFC of China under grant No.12173033, the grant from the China Manned Space Project with No. CMS-CSST-2021-B01 and CMS-CSST-2025-A03, the ``Yunnan Provincial Key Laboratory of Survey Science'' with project No. 202449CE340002 and from a ``Yunnan Provincial Top Team Projects'' with project No. 202305AT350002. C.L.W. acknowledges the support from NSFC grant No.11903082 and the grant from the China Manned Space Project with No. CMS-CSST-2021-A03.
\end{acknowledgements}

\appendix

\section{The IA setting and the projection to 2D space\label{app:IAsetting}}
\begin{figure}
    \centering
    \subfigure[\label{subfig:IAsetting}]{\plotone{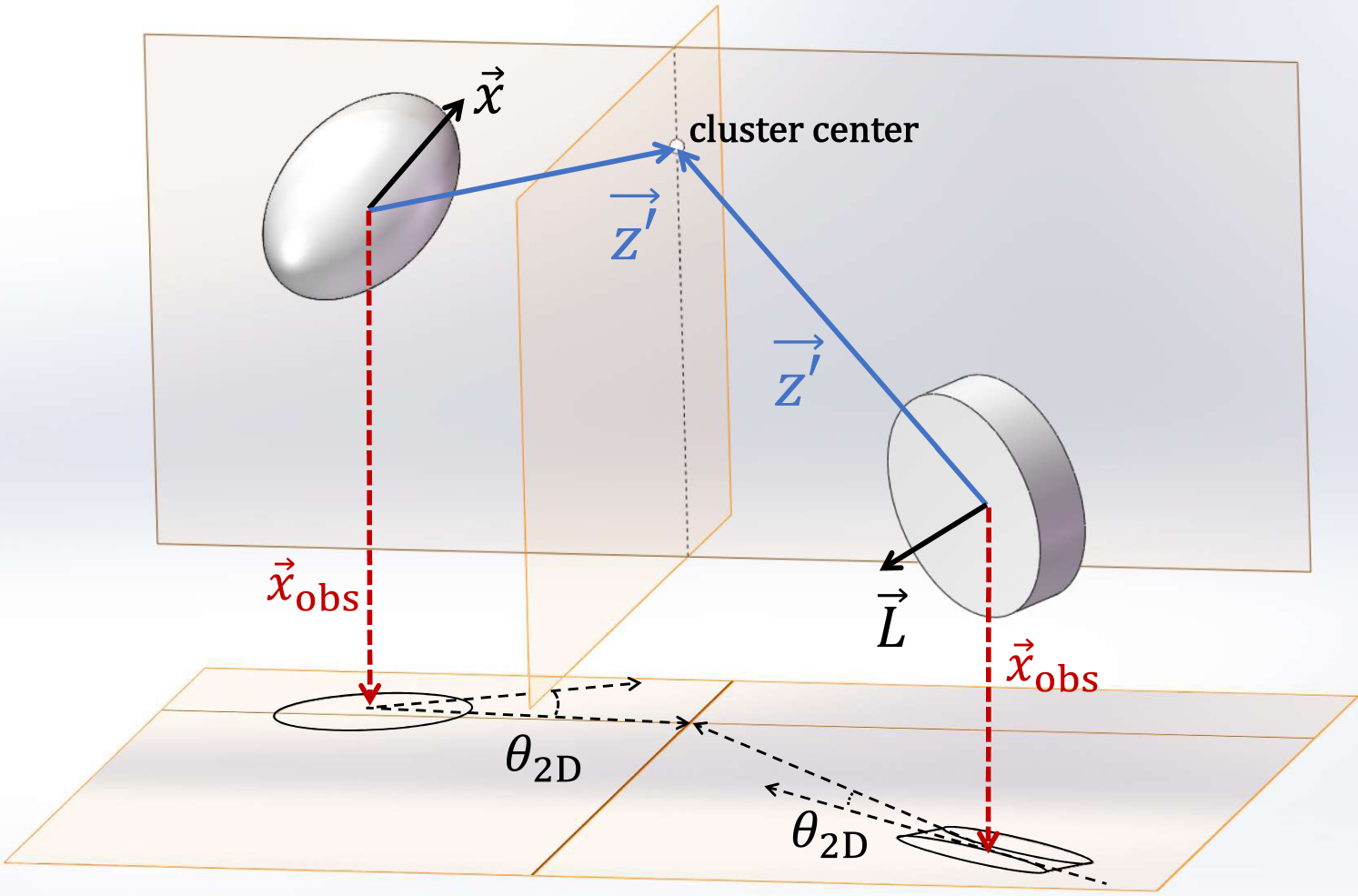}}
    \subfigure[\label{subfig:localframe}]{\plotone{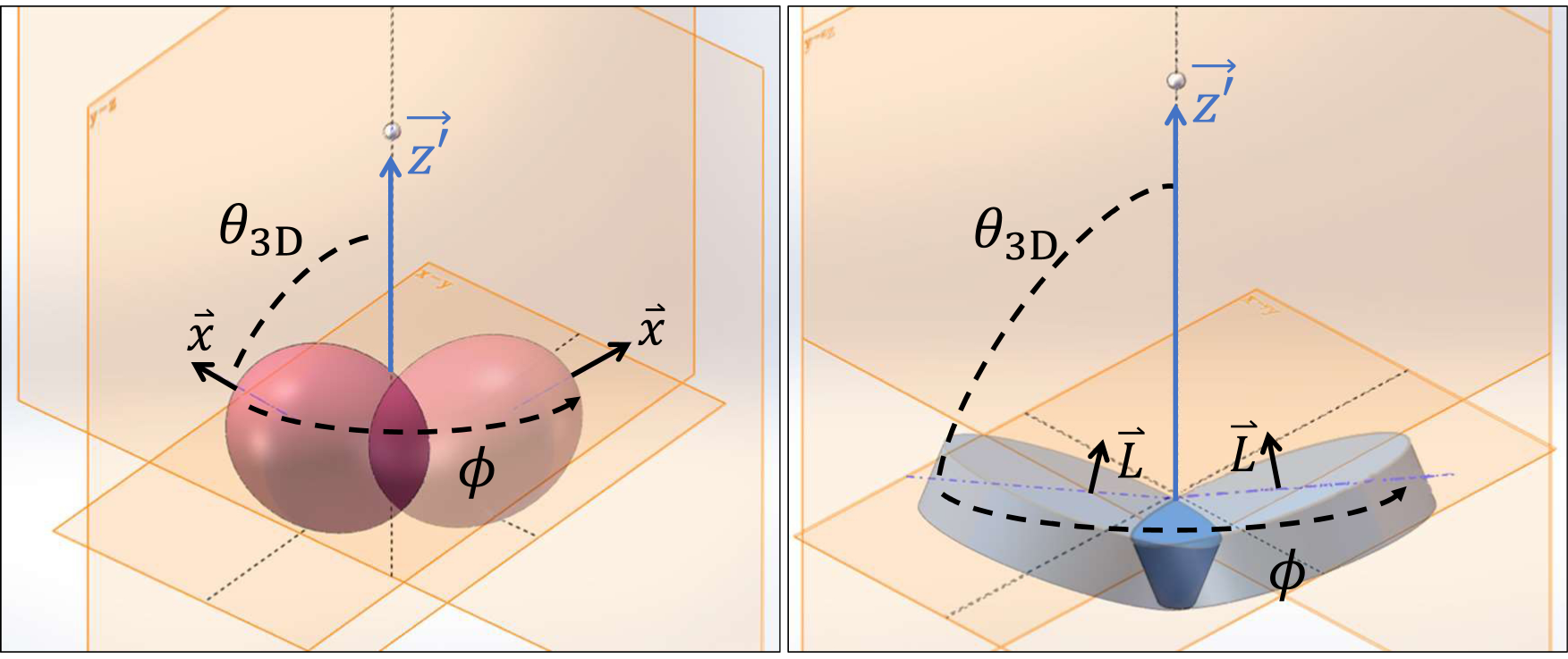}}
    \subfigure[\label{subfig:ellipproj}]{\plotone{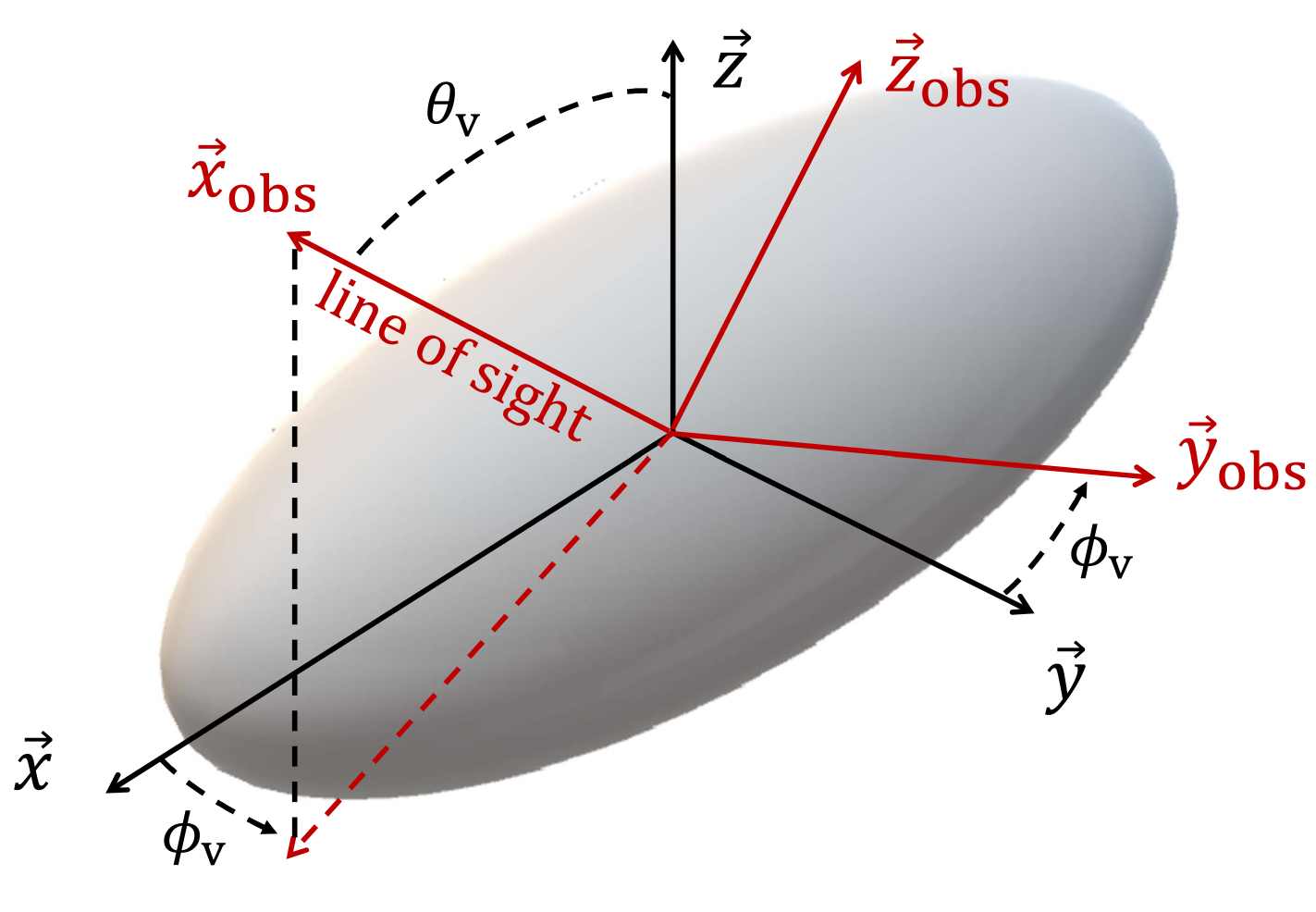}}
    \caption{Illustrations of the orientation settings of satellite galaxies. (a) A sketch of the spacial relations of satellite galaxies in 3D space, and the orientation of their projections on 2D plane; (b) An illustration of a satellite 3D IA setting $(\theta_{\rm 3D},\phi)$ with respect to the center of its host cluster with the $z^{\prime}$-direction pointing to the cluster center; 
    (c) The coordinate systems used in calculating the 2D projection for an elliptical satellite  (adapted from Fig.1 of \cite{Galletta1983}). The black, blue and red vectors present the intrinsic, the local spherical with respect to the cluster center and the observer's coordinate systems, respectively.}
    \label{fig:IAsketch}
\end{figure}

In this section we describe how to obtain the 2D satellite orientation distributions from the 3D IA model of Eq.\eqref{eq:theta_dis}. 

For an elliptical satellite galaxy, we follow the procedures of \cite{Wei} to model it in 3D as a triaxial ellipsoid given by 
\begin{equation}
    x^2+y^2/p^2+z^2/q^2=1,
\label{eq:ellipsoid}
\end{equation}
where the axial ratios satisfy $0<q\leq p\leq 1$. Then we set its orientation by rotating its long axis following the distribution of Eq.\eqref{eq:theta_dis} with respect to its host cluster center in the local spherical coordinate system. 
In Fig.\ref{subfig:IAsetting}, we illustrate the 3D setting of the satellite galaxies, the center of their host cluster and the observer's plane, where the black vector $\vec x$ indicates the long axis of the galaxy ellipsoid, and the blue vector $\vec Z^{\prime}$ shows the relative zenith direction of the satellite to the center of its host cluster. 
With respect to the cluster center, we set the 3D IA of the satellite in accord with Eq.\eqref{eq:theta_dis}, which is illustrated in Fig.\ref{subfig:localframe}. 
Finally, we calculate the 2D ellipticity of the elliptical satellite in the observer's coordinate system (red frame in Fig.\ref{subfig:ellipproj}) using the procedures of \cite{Galletta1983}. Specifically, we have 
\begin{equation}
    \begin{split}
    &\epsilon_1=\frac{1-r}{1+r}\cos(2\alpha),\\
    &\epsilon_2=\frac{1-r}{1+r}\sin(2\alpha),
    \end{split}
\label{eq:Galletta}
\end{equation}
where the apparent axis ratio of the projected ellipse $r$ and the polar angle $\alpha$ are calculated by
\begin{equation}
    \begin{split}
    &r=\sqrt{\frac{j+s-\sqrt{(j-s)^2+k^2}}{j+s+\sqrt{(j-s)^2+k^2}}},\\
    &\alpha=\arctan(k,(j-s)-\sqrt{(j-s)^2}+k^2)+\phi_{\rm v},\\
    &k=\sin(2\phi_{\rm v})\cos\theta_{\rm v}\cdot(p^2-1),\\
    &s=\cos^2\phi_{\rm v}\cdot(p^2-1)+1,\\
    &j=q^2+\cos^2\theta_{\rm v}\cdot(p^2-q^2+1-s).
    \end{split}
\label{eq:Galletta_params}
\end{equation}
The observed ellipticity of the elliptical galaxy is determined by the axial ratios $p,\ q$ and the viewing angle of the ellipsoid $(\theta_{\rm v}, \phi_{\rm v})$. The distributions of the values of $p$ and $q$ are set in accord with that measured from dark matter halos \citep{Joachimi2013,Wei}. 

For a spiral satellite galaxy, it is modeled as a disk with the ratio of the thickness to the diameter of the disk $r_d=0.25$ \citep{Wei}. The 3D orientation of the disk is determined by its spin vector $\boldsymbol{L}=(L_{\rm{los}}, L_{y_{\rm obs}}, L_{z_{\rm obs}})^T$ (see Fig.\ref{subfig:IAsetting}) where the line-of-sight direction is taken to be the direction of $\vec x_{\rm obs}$. 
Its apparent 2D axial ratio is given by \citep{Joachimi2013,Wei}
\begin{equation}
    r=\frac{|L_{\rm los}|}{|\boldsymbol{L}|}+r_d \sqrt{1-\frac{L^2_{\rm los}}{|\boldsymbol{L}|^2}},
\label{eq:diskellip}
\end{equation}
where the polar angle of the image ellipse is calculated by
\begin{equation}
    \alpha=\frac{\pi}{2}+\arctan \left(\frac{L_{z_{\rm obs}}}{L_{y_{\rm obs}}}\right).
\label{eq:diskPA}
\end{equation}

With the long axis of the projected 2D ellipse $\boldsymbol {a}$ , and the position of a satellite galaxy on the sky $\boldsymbol{\theta_{\rm s}}$, we can calculate the angle between its 2D orientation and the line connecting it to its cluster center $\boldsymbol{\theta_{\rm c}}$ on the 2D plane by
\begin{equation}
    (\sin(\theta_{\rm 2D}),0,0)^T=\frac{\boldsymbol{a}\times(\boldsymbol{\theta_{\rm c}}-\boldsymbol{\theta_{\rm s}})}{||\boldsymbol{a}||\cdot ||\boldsymbol{\theta_{\rm c}}-\boldsymbol{\theta_{\rm s}}||},
\label{eq:angdis_2d}
\end{equation}
where in the 3D observer frame, $\boldsymbol a$ can be written as $(0,a_1, a_2)^T$, so do $\boldsymbol{\theta_{\rm c}}$ and $\boldsymbol{\theta_{\rm s}}$.
The above normalized cross product has only one component in the direction of $\vec x_{obs}$ with the amplitude of $\sin(\theta_{\rm 2D})$. Taking the arcsine function of this component, we can get $\theta_{\rm 2D}$ with its sign naturally included. 

With the above, we can calculate the 2D satellite orientation distributions in accord with their 3D IA distributions, where the fractions of elliptical and spiral satellites are set in accord with our mock simulations. 

\section{examples of single-halo simulations\label{app:1haloexamples}}

\begin{figure*}
    \gridline{\fig{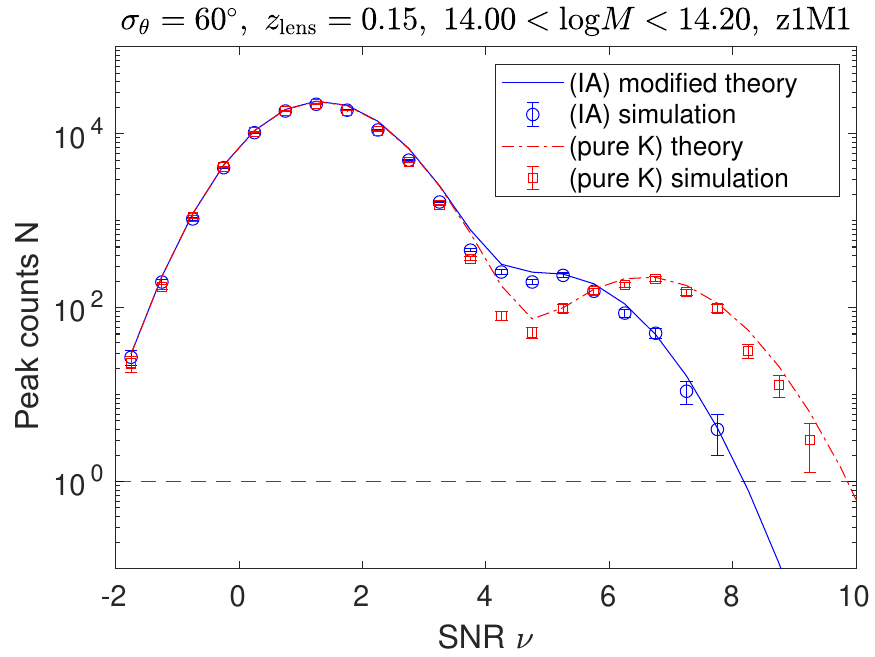}{0.32\textwidth}{}
              \fig{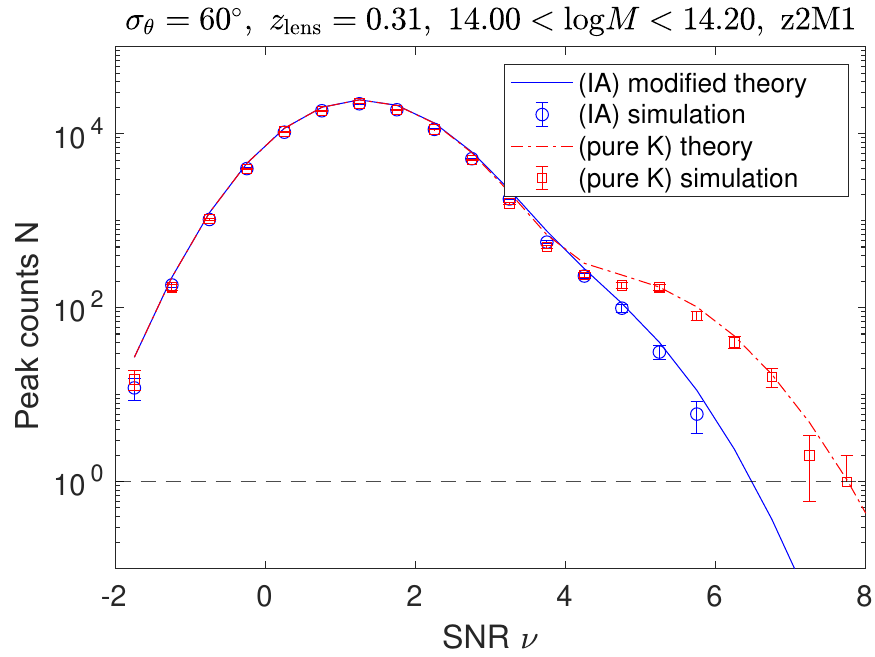}{0.32\textwidth}{}
              \fig{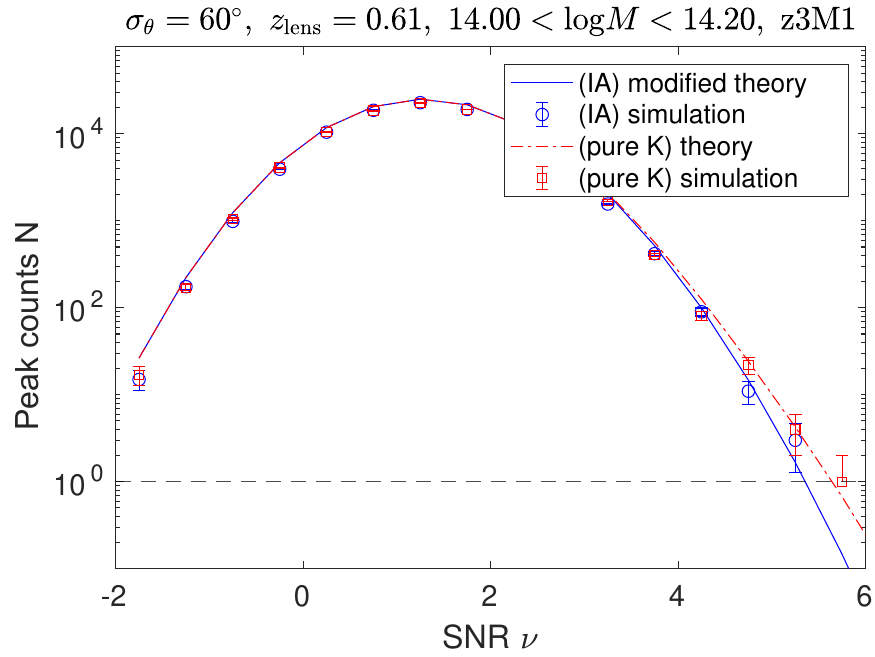}{0.32\textwidth}{}
    }
    \gridline{\fig{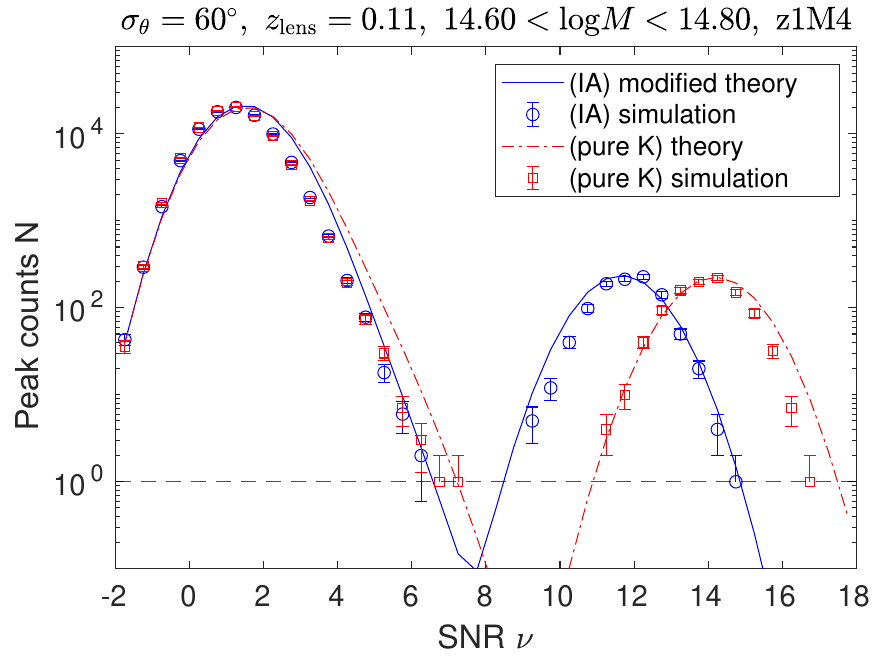}{0.32\textwidth}{}
              \fig{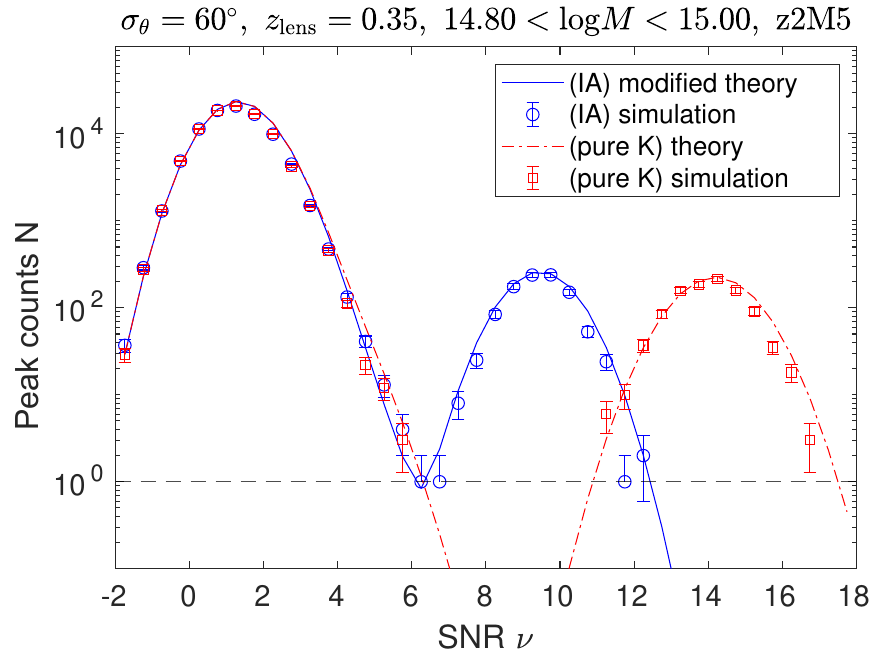}{0.32\textwidth}{}
              \fig{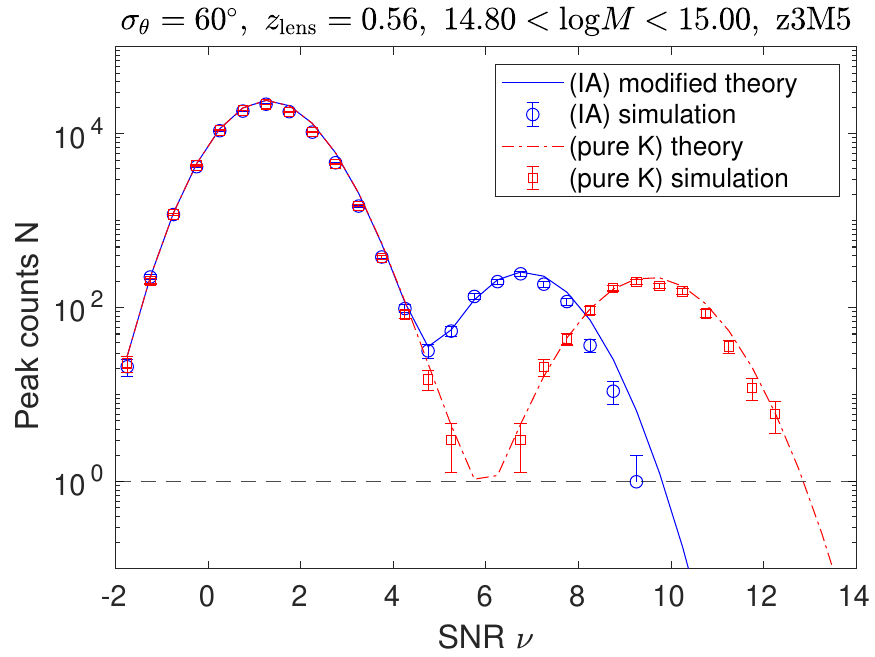}{0.32\textwidth}{}
    }
    \caption{Comparisons between the model predictions and the simulation data in the single-halo cases with different mass and redshift. Different lines and symbols are explained in the legend.
    \label{fig:onehalopeak}}
\end{figure*}

Here we present examples of the peak statistics in the cases of single-halo simulations. We take $\sigma_{\theta}=60^{\circ}$, and show the results for halos with different mass and redshift in Fig.\ref{fig:onehalopeak}. The statistics is done with $1000$ realizations in each case. The red and blue ones are for the cases without and with dilution+IA effects. The lines are the model predictions, where only one single halo with a fixed mass and redshift enters the model calculation without the mass and redshift integration in Eq.\eqref{eq:npeak_halo}. The surrounding field region in the single-halo simulation is taken into account. We note that by setting, there is no contribution from the large-scale matter projection effects, neither the large-scale IA effects here.

We can see that the effects of dilution+IA suppress the peak height from the central halo, making the blue data at high SNR part systematically shift to the left.  
Our model predictions agree with the simulation data excellently in the single-halo cases.

\bibliography{ms_revise}{}
\bibliographystyle{aasjournal}



\end{document}